\documentclass[twocolumn,aps,showpacs,amssymb,floatfix]{revtex4}
\newcommand{\be}{\begin{equation}}
\newcommand{\ee}{\end{equation}}
\newcommand{\beq}{\begin{eqnarray}}
\newcommand{\eeq}{\end{eqnarray}}
\usepackage{bm}
\usepackage{graphicx}%
\usepackage{epsf}
\usepackage{epsfig}

\begin{document}
\setcounter{figure}{\arabic{figure}}

\title{A lattice study of pentaquark states}
\author{C.~Alexandrou$^a$ and A.~Tsapalis$^{a,b}$}
\affiliation{{$^a$ Department of Physics, University of Cyprus, 
CY-1678 Nicosia, Cyprus}\\
 {$^b$ University of Athens,
Institute of Accelerating Systems and Applications, Athens, Greece}}

date{\today}%

\begin{abstract}
We present a study of the pentaquark system in quenched lattice QCD
  using  diquark-diquark and  kaon-nucleon 
local and smeared interpolating fields.
We examine the volume dependence of the spectral weights of local 
correlators 
  on lattices of size $16^3\times 32$, $24^3\times32$
and $32^3\times 64$ at $\beta=6.0$.
We find that a reliable evaluation of the
 volume dependence of the spectral weights 
requires accurate determination of the correlators at large time separations.
Our main result from the
spectral weight analysis 
in the pentaquark system 
is that within our variational basis and statistics we can not 
exclude  a pentaquark resonance.
However our data also do not allow a clear identification of a pentaquark state
 since only the spectral weights of the lowest state can be
determined to sufficient accuracy to test for volume dependence.
 In the negative parity channel
the mass extracted for this state is very close to the KN 
threshold whereas in the positive parity
channel is about 60\% above.
\end{abstract}


\maketitle


\section{Introduction}

Experimental 
searches
for  a pentaquark state near the KN threshold,
reported in recent experiments~\cite{LEPS,experiment},  
are under way  world wide.
The accumulation of evidence from low energy experiments for
the existence of this state~\cite{positive} combined with the negative
results obtained in high energy experiments~\cite{negative} pose deep questions
as to its nature and its production mechanism. Further doubts as 
to the existence of the $\Theta^+$ have been  raised since
 the resent report of the CLAS collaboration for lack of evidence for
a resonance state 
from a dedicated high statistics proton experiment 
~\cite{negative2}.
The $\Theta^+$  was predicted theoretically in the chiral soliton 
model~\cite{soliton} as 
an exotic baryon state with an unusually  narrow width.  
The possible existence of such a state has
 raised interesting questions on what its structure 
should be in order to account for
its narrow width.
A number of phenomenological models have been put forward to explain 
its stability 
such as special flux tube configurations~\cite{nussinov,Lipkin,diamond,AK} 
and diquark formation~\cite{Jaffe}.
 
Several studies  in lattice QCD have looked
for a pentaquark state in order to
determine its mass and parity but no consensus
has been reached yet with some groups
finding a bound state with mass close 
to the experimental value~\cite{Wuppertal,Chiu,Sasaki,latt04,japan}
and others the KN scattering state~\cite{Mathur,Ishii,Adelaide}.  
One main difference between these groups has been the
interpolating field  used to create this state~\cite{summary}. 
Since its structure
is unknown, optimizing the interpolating field that one uses is a  difficult
task. However, for reasonable interpolating fields, the results 
 should not be
dependent on the interpolating field. In this work we use 
 two of the most popular choices: an interpolating field motivated by
the diquark-diquark picture~\cite{Jaffe} and an   interpolating
field motivated  by the  kaon-nucleon structure~\cite{Lipkin}.
We note that diquark formation was shown to be important
in the context of the static pentaquark potential where  arrangements 
of quarks that allow diquark formation are found to be energetically favored
 giving rise to a  potential that is proportional to the minimal flux
tube connecting the five quarks~\cite{AK,latt04,pentaq}. 
Using these  interpolating fields as a basis we construct a $2\times2 $ 
mass correlation matrix
 and determine the optimal linear combination that yields maximum overlap with
the  ground state of the system.
Another way of enhancing the  ground state
by suppressing excited state contributions is smearing.
In addition to using local interpolating
fields, we also use smeared fields by applying
  gauge invariant smearing to the quark fields 
 used.
We find  that the 
 value we obtain for the  mass
is independent of which of these various interpolating fields we use.
Since both interpolating fields
 have an overlap with the KN scattering state 
 we expect, that being the lowest state, 
  it will determine the 
 large time dependence of both correlators.
If a pentaquark bound state 
exists about 100 MeV higher than the KN
threshold we expect it to dominate the time dependence of the
correlator up to 
time separations of about   $\sim 10$~GeV$^{-1}$. 
Therefore the main difficulty is to identify unambiguously the
resonance in an intermediate time range before
 the nearby KN scattering state  dominates.
If our interpolating fields have good overlap with the KN
scattering state and the $\Theta^+$, then diagonalization of the
mass  correlation matrix should give as lowest eigenvalues these
two states. 
This is our motivation for choosing the KN interpolating field, which
is expected to have large overlap with the KN scattering states
and the diquark-diquark field expected to have large enough
overlap with the pentaquark state.
To distinguish a single particle state from a
scattering state the tools that we have at our disposal are  the volume
dependence of the energy and spectral weights of the state provided
we can isolate it with sufficient accuracy.
 The energy of a scattering state on a lattice,
 with the exception of  s-wave scattering states,  is 
volume dependent. This is because, 
in the center of mass frame, the two particles  have
non-zero relative momentum, which on a finite lattice, depends on the 
spatial size of the lattice. This means that if a scattering state
with non-zero momentum dominates the correlator
then 
its energy will
depend on the spatial length of the lattice.  
The other criterion that can be used is
 the scaling of the spectral weights with the spatial volume.
For a  two-particle scattering 
state the spectral weights are expected to be
 inversely proportional to the spatial volume 
 whereas for a resonance there should be no volume dependence.
The spectral decomposition of the correlator
was given
by L\"uscher~\cite{Luscher} who first noted
that a weakly interacting two-particle state well below resonance
energy should contribute to the correlator with 
an amplitude that scales inversely proportional to the volume.
The scaling of the spectral weights
as a probe for a two-particle resonance scattering state
was first  used in a lattice study
of the pentaquark system by Mathur {\it et al.}~\cite{Mathur}.
In order to check how reliably we can extract the
scaling of spectral weights in practice  we study the two pion
system in the isospin I=2 channel for which no 
low energy
resonances are present. This study is done on 
 three lattices of size $16^3\times 32$, $24^3\times 32$ and
$32^3\times 64$ at $\beta=6.0$, which are 
the same lattices used in the study of the pentaquark system. 
We consider four different two pion interpolating fields 
and determine accurately
the two lowest energy eigenstates. What we find is that the value
of spectral weights extracted stabilizes
only at large enough time
separations and only then
  the ratio of spectral weights
reaches the expected value. Therefore to check for
volume dependence one requires
very accurate  data so that the extracted ratio of  spectral weights  
is precise enough. 
For the two-pion system, given our statistics,
this can only be achieved for the lowest eigenstate.
How large contributions from higher relative 
momentum states are can be assessed by
 explicitly projecting  to 
 zero relative momentum. This is done   for  the lattice 
of size $16^3\times 32$ and  the results
obtained are compared
to those without explicit projection to zero relative momentum.
We note that such a comparison has not been carried out in previous lattice
studies of the pentaquark system.
We confirm
that diagonalization of the correlation matrix without explicit
projection  yields a lowest energy eigenvalue
that 
contains contributions from states with higher momenta
and only becomes a pure s-wave scattering state at large time separations.
This means that in order to obtain the lowest 
scattering state
one must allow for larger time separation as the spatial extent of
the lattice increases. As we will demonstrate
for a lattice of spatial size of 3~fm one has to go beyond a time interval
of  $30a \sim 15 GeV^{-1}$ to obtain the
lowest scattering state. 
This is why we use Dirichlet boundary conditions (b. c.) in
the temporal direction.
In the study of  the two-pion system we use the heaviest pion
mass considered in this work,
namely $\kappa_l=0.153$,  to have the smallest 
statistical errors.
We compare the scaling behavior  of spectral weights 
in the two-pion  and 
pentaquark system using the same value of $\kappa_l$.  
Our main conclusion is that within the accuracy that
 the spectral weights can be determined from our data we  can not exclude 
 a pentaquark resonance.
 In the negative
parity channel within our variational basis 
we only obtain the lowest eigenstate accurately enough to
be able to perform the scaling analysis of the spectral weights.
Because of this we are unable to draw a definite conclusion regarding
the existence of the $\Theta^+$. For the positive parity channel 
we obtain two eigenstates very close in energy which
at the two heaviest light quark masses have an energy gap of about 100 MeV
 independent of the spatial volume.  
As the quark mass gets lighter it becomes more difficult to obtain
an accurate determination of the energy gap. Although the energy gap
that we find is of the right order of magnitude
the mass for this state is high:  Measuring the  mass 
in the positive parity channel  for five values of 
light quark masses
on our largest lattice using a smeared  
diquark-diquark interpolating field for the source
and a local one for the sink 
and extrapolating  to the chiral limit we obtain a value of
$1.65 \pm 0.09$ times  the mass of the non-interacting 
KN state.
This value is 
too high to be identified with the
$\Theta^+(1540)$.
 The values of the light quark
mass that we use in carrying out the
extrapolation correspond to pion masses in the range of about 900-420~MeV.

\section{Lattice techniques}
In this work we consider two interpolating fields motivated by recent
proposals on the possible structure of the $\Theta^+$ state. The first
is based on the idea of  diquark
formation~\cite{Jaffe} and the other on a diquark-triquark  
structure~\cite{Lipkin}.
 Both have been used in previous lattice 
studies
but a correlation matrix analysis in the manner
considered here has not been presented.
We will refer to the first as diquark-diquark interpolating field. 
It is given by
\be
{\cal J}_{DD} =\epsilon^{abc}\epsilon^{aef}\epsilon^{bgh} \> C\bar{s}_c^T
\left(u_e^T C d_f\right) \left(u_g^T C \gamma_5 d_h\right)
\label{diquark-field}
\ee
where $C=\gamma_0\gamma_2$ is the charge conjugation operator.
It was first proposed and used in the study of the pentaquark system
on the lattice
by Sasaki~\cite{Sasaki}.
After performing the antisymmetric tensor contraction, we generalize the 
resulting expression to obtain an isospin I=0 and I=1 interpolating fields: 
\be
{\cal J}_{DD} =\epsilon^{abc} \> \left(u_a^T C\gamma_5 d_b\right)
\left[ (u_c^T C d_e) \mp  (u_e^T C d_c) \right] \> C\bar{s}_e^T \quad.
\ee
The minus sign corresponds to isospin I=0 and the plus sign to isospin I=1 
respectively.
The second field, which we will refer to as 
the KN-interpolating field is given by
\be
{\cal J}_{NK} =\epsilon^{abc} \> \left(u_a^T C\gamma_5 d_b\right)
\left[ u_c (\bar{s} \gamma_5 d)\mp  d_c (\bar{s}\gamma_5 u)\right] \quad,
\label{KN-field}
\ee
where the minus sign corresponds to the isoscalar and the plus
to the isovector. 
Although we have used both the isoscalar and the 
isovector interpolating fields, in this work we will only discuss the results
in the isospin zero channel.  

As presented, the interpolating fields have opposite parities with
${\cal J}_{DD}$ having positive parity and ${\cal J}_{KN}$ negative. The
parity of these fields can be flipped by multiplying with $\gamma_5$.
The two point correlator can couple both to positive and negative parity
states and  the propagators for the positive and negative parity interpolating
fields are given, respectively, by~\cite{Sasaki2}
\beq
G_{+}(t) &= &\frac{(1+\gamma_0)}{2} g(t) + \frac{(1-\gamma_0)}{2} g(-t)\\ \nonumber
G_{-}(t) &= &\frac{(1-\gamma_0)}{2} g(t) - \frac{(1+\gamma_0)}{2} g(-t)\\ \nonumber
g(t)& = &\theta(t)C_{+} e^{-m^{+} t} + \theta(-t) C_{-} e^{m^{-}t} \quad,
\label{G}
\eeq
where $m^{+}$ ($m^{-}$) is the mass of the positive (negative) parity state.
Employing Dirichlet boundary conditions means that only the terms with $t>0$
contribute and therefore the positive and negative
 parity states correspond to the upper
 and lower Dirac  components of $G_+(t)$ or the lower and upper components
of $G_-(t)$ respectively.  
For the two smaller volumes we use only Dirichlet b.c. 
and local
interpolating fields since the main purpose of the evaluation on the
smaller lattices is to examine  the volume dependence of the spectral 
weights in the correlators.
 In order to perform our
volume studies we also use Dirichlet b.c. for the local interpolating field 
on our largest volume. However the $32^3\times 64$ lattice
 has a time extent large enough so that backwards moving
quarks are suppressed and anti-periodic boundary conditions can also be 
used without affecting the identification of the parity.
Therefore on the large
volume we opt for anti-periodic b.c. when using smeared sources.
It is well known that smearing  improves the overlap
of the interpolating field with the ground state
since it produces an extended source of size typical 
to that of physical hadrons.
 We perform gauge
invariant smearing~\cite{smear} by replacing a local quark field $u(x)$ 
appearing
in the interpolating fields by a smeared one, $\tilde{u}(x)$, obtained by
\be
\tilde{u}({\bf x},t)= \sum_{\bf y} \Phi({\bf x},{\bf y};U(t)) u({\bf y},t) \quad.
\label{smearing}
\ee
The gauge invariant smearing function  $\Phi({\bf x},{\bf y};U(t)) $ 
is given by
\be
 \Phi({\bf x},{\bf y};U(t)) =(1+\alpha H)^n({\bf x},{\bf y};U(t))
\label{smaearing function}
\ee
where the hopping matrix $H$ is defined by
\be
 H({\bf x},{\bf y};U(t))=\sum_{j=1}^3 
         \left[ U_j({\bf x},t) \delta_{{\bf x, y}-\hat{j}}
              + U_j^{\dagger}({\bf x}-\hat{j},t) \delta_{{\bf x, y}+\hat{j}}
             \right]
 \quad.
 \label{hopping}
\ee
The parameters $\alpha=4$ and $n=50$ are optimized to approximately
reproduce the root mean square radius of the nucleon
in the quenched theory at $\beta=6.0$ where all the
computations are carried out. 

Having M interpolating fields we can construct an $M\times M$ mass correlation
matrix
\be
{\cal C}_{i;j}(t)=\int d^3x <0|{\cal{J}}_i({\bf x},t){\cal{J}}_j^\dagger({\bf 0},0)|0>
\ee
where in the case of the pentaquark system the indices $i,j=DD$ or $KN$. 
 For non-vanishing results we must use interpolating
fields of the same parity. This can be easily achieved by multiplying
for example ${\cal J}_{DD}$ with $\gamma_5$.
Our variational analysis is performed in two ways:\\
1. We solve the generalized
eigenvalue equation
\be
{\cal C}(t_1)v_n(t_1)=\lambda_n(t_1, t_0){\cal C}(t_0)v_n(t_1) \quad.
\label{eigenvalue equation}
\ee 
For a large 
time separation $t_1-t_0$ 
the eigenvalues are
given by
$\lambda_n(t_1,t_0)=\exp(-E_n(t_1-t_0))$, $n=0,1,..,M-1$,  
yielding the energies $E_n$ of the M lowest states~\cite{variational} with
total momentum zero.
The energy of the M$^{th}$ eigenstate is usually poorly determined since it
has contributions from all the higher excited states, except  when 
our variational basis contains interpolating fields with a sizable
overlap with all M lowest states.
In this analysis we take $t_0/a=3-6$ with $a$  the lattice spacing and 
check that  the values that we find for $E_n$ do not 
change as we vary $t_0$.\\
2. We first diagonalize the correlation matrix ${\cal C}(t_0)$
\be
{\cal C}(t_0)v_n(t_0)=\lambda_n(t_0)v_n(t_0)
\label{eig0}
\ee
to determine the eigenvectors $v(t_0)$ taking $t_0/a=1$. We use these 
eigenvectors to project to the space spanned by the
 N largest eigenvalues $\lambda_n(t_0)$
\be
{\cal C}_{ij}^N(t)=(v_i,C(t)v_j), \hspace*{0.5cm} i,j=0,..,N-1
\label{project correlation}
\ee
and solve the generalized eigenvector equation given in 
Eq.~\ref{eigenvalue equation} but for the projected correlation matrix
 ${\cal C}_{ij}^N$ instead of
${\cal C}(t)$. We denote the
resulting eigenvalues by $\Lambda_n$.
The eigenvectors, ${\cal V}_n$, that we find  determine
the best linear combination $\sum_n {\cal V}_n {\cal J}_n$
 that has maximum overlap with the N
lowest  eigenstates. We will refer to this linear  combination as the
optimal interpolating field, ${\cal J}_{optimal}$. The optimal correlator
can be obtained by projecting $C^N$:
$({\cal V}_i,C^N {\cal V}_j)$ which
yields the same  energies as those extracted from the
eigenvalues $\Lambda_n$. We check that the eigenvalues,
${\Lambda}_n$
that we obtain by diagonalizing ${\cal C}_{ij}^N$
are in agreement with $\lambda_n$.
 
For the
pentaquark system, in addition,
 to ${\cal C}_{DD;KN}$ constructed using the local
interpolating fields  ${\cal J}_{DD}$ and ${\cal J}_{KN}$ as a basis, 
we also consider
  $\tilde{\cal C}_{DD;KN}$ constructed
using smeared  fields  $\tilde{\cal J}_{DD}$ 
and $\tilde{\cal J}_{KN}$ for the source but keeping
the sink local and vice versa. 
The eigenvalues, $\tilde{\Lambda}_n$,
 extracted from this correlation matrix 
should yield, for large enough time separations, the same energies
as those extracted from $\Lambda_n$.
We
check for consistency at the two heaviest light quark masses, 
namely $\kappa_l=0.153$ and $\kappa_l=0.155$.
  
The contractions needed for the computation of the pentaquark matrix
elements are optimized by  doing all the Dirac contractions explicitly 
before summing over color. For  example, using the ${\cal J}_{DD}$ 
interpolating
 field as source and sink, the correlator is given by
\begin{widetext}
\beq
{\cal C}_{DD;DD}(t) =\sum_{\bf x} \epsilon^{a^\prime b^\prime c^\prime} 
\epsilon^{abc} \left[-C\gamma_5 S^*(x) C\gamma_5 \right]^{f^\prime f}_{\mu^\prime \mu^\prime}
\biggl\{\biggl[ &\>& 
{\rm Tr}\left(C\gamma_5 D^{c^\prime c}(x)C\gamma_5 U^{b^\prime b \>T}(x)\right)
{\rm Tr}\left(C D^{f^\prime f}(x) C U^{a^\prime a \>T}(x)\right)\nonumber \\
&-& {\rm Tr}\left(C\gamma_5 D^{c^\prime f}(x) C U^{b^\prime b \>T}(x) \right)
{\rm Tr}\left(C D^{f^\prime c}(x) C\gamma_5 U^{a^\prime a \>T}(x)\right)
\nonumber \\
&+& {\rm Tr}\left(C\gamma_5 D^{c^\prime c}(x) C\gamma_5 U^{a^\prime a \> T}(x) 
           C D^{f^\prime f}(x) C U^{b\prime b \> T}(x)\right) \nonumber \\
&-& {\rm Tr}\left(C\gamma_5 D^{c^\prime f}(x) C U^{a^\prime a \> T}(x) 
           C D^{f^\prime c}(x) C\gamma_5 U^{b\prime b \> T}(x)\right) 
 \> \biggr ]\nonumber \\
&\>& \hspace*{-1cm} 
+\biggl[ \> a^\prime \leftrightarrow f^\prime;a\leftrightarrow f\>\biggr]
- \biggl[ \> a^\prime \leftrightarrow f^\prime \>\biggr]
- \biggl[ \> a \leftrightarrow f \> \biggr] \biggr\} \quad,
\eeq
\end{widetext}
where $S(x), U(x)$ and $D(x)$ denote the strange, up and down quark 
propagators, Latin letters  denote color indices, Greek letters 
Dirac indices and the transpose acts only on Dirac indices.
In each of the terms we perform the Dirac algebra separately and store the 
result in an array e.g. in the first term having two
traces we construct the traces and then we multiply them and
sum over color indices.    This reduces the time needed by more than an order of magnitude 
to about twice the time
 one needs for the nucleon correlator.
Such grouping of terms is common in many
lattice studies of matrix elements  as for example 
in Refs.~\cite{Leinweber,Mathur,Adelaide}.

\section{Results}

The lattices and values of the hopping
parameter, $\kappa_l$, that we use for the light quarks
are listed in Table~\ref{table:parameters}
where we
also give the mass of the pion, the kaon and the nucleon at
each value of $\kappa_l$ as well as the value obtained by linear
extrapolation to the chiral limit using the form
\be
m_H^2=a + b m_{\pi}^2  \quad.
 \label{chiral}
\ee
In Eq.~\ref{chiral}
 $m_H$ denotes the mass of any of the hadrons  considered in this work.
In Table~\ref{table:parameters} we also include the mass of the negative parity
baryon $N^*$ at the values of $\kappa_l$ where we could identify a plateau.
For the strange quark we take
$\kappa_s=0.155$, which produces a mass for the $\phi$ meson of $0.421(3)$  in lattice units.
Using  the mass of the nucleon  in the chiral limit 
given in Table~\ref{table:parameters}  
 we find that
the ratio of the mass of the $\phi$ meson to the mass of the nucleon
at the chiral limit is $m_{\phi}/m_N=1.002 \pm 0.025$, which is  very close to
the physical ratio of $1.087$ verifying that the value of $\kappa_s=0.155$ used
is very close (within 10\%) to
the physical strange quark mass. We analyzed 202 configurations 
for the $16^3\times 32$ lattice~\cite{connection}
and 100 configurations for each of  the $24^3\times 32$ and $32^3\times 64$ lattices.

\begin{table}
\caption{In the first column we give the lattice volume 
and  in the second column the value of the hopping parameter, $\kappa_l$, for
light quarks.
 In the third, fourth, fifth and sixth columns we give
 the pion mass, the kaon mass, the nucleon mass and the $N^*$ mass in lattice 
units. For $\kappa_l=0.1558$ and 0.1562 the mass of $N^*$ is not given
since no clear plateaus could be identified. }
\label{table:parameters}
\begin{tabular}{|c|c|c|c|c|c|}
\hline
\multicolumn{1}{|c|}{volume } &
\multicolumn{1}{ c|}{$\kappa_l$ } &
\multicolumn{1}{ c|}{$m_\pi$ } &
\multicolumn{1}{ c|}{$m_K$ } &
\multicolumn{1}{ c|}{$m_N$ } &
\multicolumn{1}{ c|}{$m_{N^*}$ }\\
\hline
\multicolumn{6}{|c|}{local source and sink using Dirichlet 
b.c. and ${\cal J}_{DD}$, ${\cal J}_{KN}$}
 \\ \hline
 $16^3\times 32$ &  0.153 & 0.422(2) &  0.363(2) & 0.787(8)&  1.039(56)\\
                 &  0.155 & 0.295(3) &  0.295(3) & 0.623(14)&  0.895(86)\\
 $24^3\times 32$ &  0.153 & 0.426(3) &  0.369(2) & 0.790(10)& 1.026(42)\\
                 &  0.155 & 0.302(3) &  0.302(2) & 0.644(12)&  0.877(57) \\
 $32^3\times 64$ &  0.153 & 0.420(2) &  0.361(2) & 0.788(6) & 0.987(33)\\
                 &  0.155 & 0.294(2) &  0.294(2) & 0.633(7) & 0.849(46)\\
\hline
\multicolumn{6}{|c|}{smeared source and local sink using 
anti-periodic b.c. and ${\cal J}_{DD}$}
 \\\hline
$32^3\times 64$  &  0.153  & 0.418(2) & 0.358(2) & 0.778(6) & 0.990(30)\\
                 &  0.1550 & 0.292(2) & 0.292(2) & 0.625(7) & 0.802(44)\\
                 &  0.1554 & 0.262(2) &  0.277(2)& 0.590(9) & 0.748(49)\\
                 &  0.1558 & 0.229(2) &  0.262(2)& 0.553(9) & \\
                 &  0.1562 & 0.192(2) &  0.246(2)& 0.513(10)& \\
    &  $\kappa_c=$0.1571   & 0.       &  0.207(5)& 0.420(10)& \\
\hline
\multicolumn{6}{|c|}{smeared source and  sink using 
anti-periodic b.c. and ${\cal J}_{DD}$}
 \\\hline
$32^3\times 64$ &  0.153 & 0.418(2) & 0.360(3) & 0.789(9)&  0.973(39) \\
\hline
\end{tabular}
\end{table}

\subsection{The nucleon sector}
\begin{figure}
\epsfxsize=8truecm
\epsfysize=10.truecm
\mbox{\epsfbox{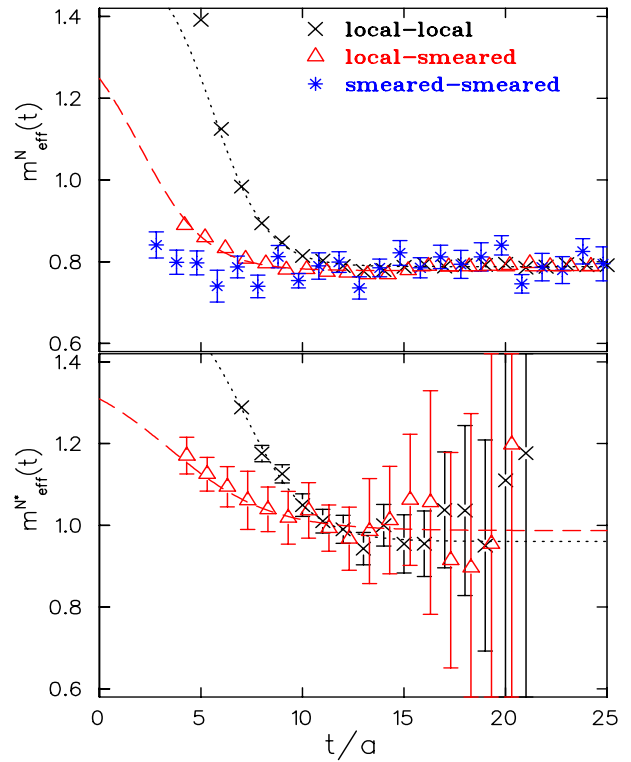}}
\caption{The upper graph shows
the effective mass for the nucleon (positive parity)  and the lower graph
for the $N^*$ (negative parity) at $\kappa_l=0.153$. The crosses
show the results using the local-local correlator, $C(t)$, and Dirichlet b.c. 
and the open triangles
using the local-smeared correlator, $\tilde{C}(t)$ and anti-periodic b.c.
shifted to the right for clarity. 
In the case of the nucleon we also show 
results obtained using the smeared-smeared correlator
$\widehat{C}(t)$ (asterisks)  shifted to the
left for clarity. 
The dotted and dashed lines are  fits to the effective masses
obtained using  $C(t)$ and $\tilde{C}(t)$ respectively
assuming two state dominance.}
\label{fig:nucl_meff}
\end{figure}

In order to assess our methods of analysis we first examine the results
obtained for the nucleon using the standard interpolating field
\be  
{\cal J}_N(x)=\epsilon^{abc}(u^T_a(x)C\gamma_5 d_b(x))u_c(x),
\label{nucleon field}
\ee
which can be local as given in Eq.~\ref{nucleon field} 
or  constructed from smeared fields 
$\tilde{u}$ and $\tilde{d}$ instead of $u$ and $d$.
We denote the latter by $\tilde{{\cal J}}_N$. 
In evaluating the nucleon correlator we can use ${\cal J}_N$ for the
source and sink or use  
$\tilde{{\cal J}}_N^\dagger$ for the source and ${\cal J}_N$ 
for the sink.
In general, denoting with $H$  the appropriate hadronic state,
 the correlator
$
C(t)=\int d^3x <0| {\cal J}_H(x) {\cal J}_H^\dagger(0)|0> 
$
 computed using
local interpolating fields 
for the source and the sink is  referred to as the local-local 
correlator, whereas  the correlator $\tilde{C}(t)$
computed using a local (smeared) field for the source and 
smeared (local)  for the sink
is referred to as local-smeared correlator. Finally
the correlator, $\widehat{C}(t)$, computed using  
 smeared fields  for both the source and the sink is referred
to as smeared-smeared correlator.
In Fig.~\ref{fig:nucl_meff} we show  results for the nucleon 
effective mass defined by
\be
m_{\rm eff}(t) = -\log\frac{C(t)}{C(t-1)} 
\ee
on the lattice of size $32^3\times 64$  with
Dirichlet boundary conditions. On same figure
we also show results obtained from   $\tilde{C}(t)$
with anti-periodic boundary conditions in the time direction. 
As can be seen smearing 
improves the overlap with the nucleon ground state resulting in an
earlier plateau. All the errors shown on our figures and given in the tables
are statistical  and they are determined using a jackknife analysis.

\begin{figure}

\epsfxsize=9truecm
\epsfysize=10.truecm
\mbox{\epsfbox{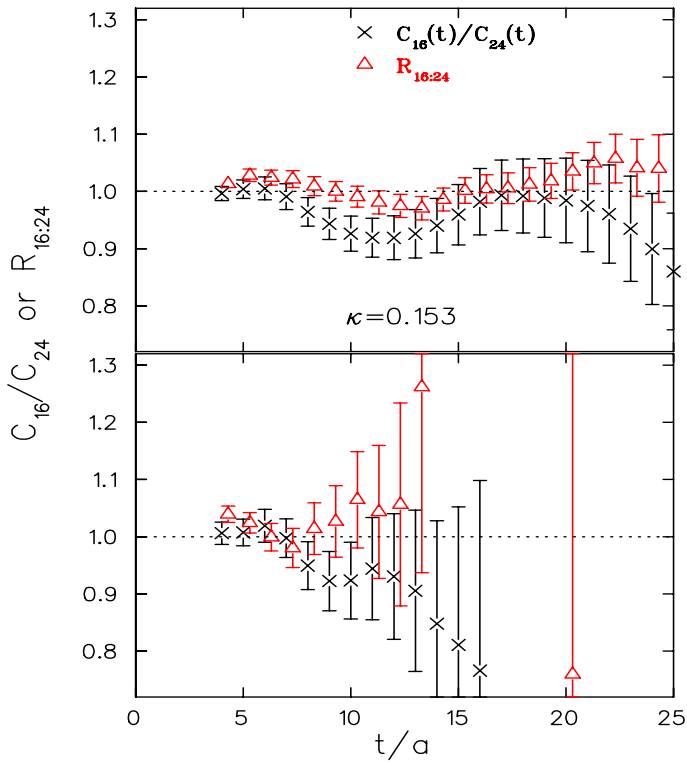}}
\caption{We show the ratio $C_{16}(t)/C_{24}(t)$ (crosses) and the
ratio $R^H_{16:24}$ (open triangles with a small shift to the right for
clarity) as a function of the time separation in lattice units.
 The upper graph shows  results for 
 the nucleon and the lower for $N^*$
at $\kappa_l=0.153$.}
\label{fig:nucl_zratio}
\end{figure} 
 We fit the effective mass assuming two state dominance
to the form
\be
m_{\rm eff}(t)
   = m_0 - \log\frac{1+z e^{-\Delta m t}}{1+z e^{-\Delta m (t-1)}}
\label{two exponential fit}
\ee
where $m_0$ is the mass of the ground state,
$\Delta m=m_1-m_0$ is the mass  gap between
the ground state and the first excited state  
and $z$ is the ratio of  the overlap of the interpolating field with
the first excited state as compared to the ground state.
For the nucleon sector $m_0$ gives the mass of the nucleon and
$\Delta m$ the mass gap between the nucleon and the N(1440) state.
The values of $m_0$ and $\Delta m$ 
determined from fitting the effective mass derived either from  
$C(t)$ or  $\tilde{C}(t)$
are in agreement. The errors on the
 mass gap $\Delta m$ however are large not allowing an accurate determination 
of the mass of the excited state.
The fact that the ground state overlap is
enhanced  as compared to that of the first excited 
state  when smearing is used, is confirmed by the decrease in the
 value of $z$ by a factor of about twenty.
Smearing 
both source and sink tends to suppress even further 
the excited state contributions
and decorrelates 
the results at successive time slices. 
The mass extracted by fitting the 
smeared-smeared correlator, $\widehat{C}(t)$, to a single exponential 
is $m_N=0.789(9)$ consistent
with that extracted from the local-smeared correlator. However, as can
be seen in Fig.~\ref{fig:nucl_meff},  the
statistical error on the smeared-smeared results
at each time slice
is larger as compared to the errors of the local-smeared results 
due to the gauge noise introduced at the sink.
Such errors will make it difficult to distinguish two closely lying
states and therefore in the case of the pentaquark system we will mostly use
local-local and local-smeared correlators.

An important test that distinguishes particle states from scattering states 
is the scaling of the spectral weights of 
local correlators with the spatial volume of
the lattice~\cite{Luscher,Mathur}. Expanding the correlator
computed on a lattice of spatial size $L$
in terms of states with the same quantum numbers as the
interpolating field  one obtains 
\be
C_{L}(t) = \sum_{\alpha} W^{\alpha}_L e^{-E_L^\alpha t} \quad.
\label{correlator2}
\ee
For a single particle state
the weights $W^{\alpha}_L$ are of order one whereas for a scattering
state consisting of two weakly interacting particles well below the resonance 
are of order $1/L^3$.
Furthermore, for a single particle state, $E_L^{\alpha}$ determines
the mass of the state and should be volume independent 
for large enough volumes.
For a scattering state that is not an s-wave, on the other hand,
$E_L^{\alpha}$ is volume
dependent since it involves the
relative momentum of the two particles, a fact which on
the lattice means that each particle carries equal 
momentum in units of  $2\pi /L$.
In the case of the nucleon we know
that we have a single particle state
and we can check  how reliably we can extract these weights.
 In Fig.~\ref{fig:nucl_zratio} we plot the ratio 
of the local-local correlator $ C_{16}(t)$ to $ C_{24}(t)$
(computed on the lattices of size $16^3$ and $24^3$ respectively)
for the nucleon  and
the $N^*$. 
In a time interval where a single state
dominates (plateau region), assuming its mass is volume independent,
 the ratio of correlators is equal to the ratio of weights $W_{16}/W_{24}$.
This ratio should be one for a single particle state.
As can be seen the ratio of correlators
is approximately one.
In the same figure we also show  the ratio 
\be
R^H_{L_1:L_2}\equiv 
\frac{C_{L_1}(t) e^{m_{\rm eff,L_1}(t)}}{C_{L_2}(t) e^{m_{\rm eff,L_2}(t)}} \quad, 
\label{scaling}
\ee
where with $H$ we denote the nucleon or the $N^*$ 
and we have taken $L_1=16$ and $L_2=24$.
This ratio 
corrects small finite size effects on the mass and in the plateau
region is again equal to  $W_{L_1}/W_{L_2}$. As can be seen $R^H_{L_1:L_2}$
indeed improves the signal, in particular for the nucleon,
 giving better agreement
with unity, which is  the expected result.

\subsection{The two pion sector}
\begin{figure}
\epsfxsize=8truecm
\epsfysize=11.truecm
\mbox{\epsfbox{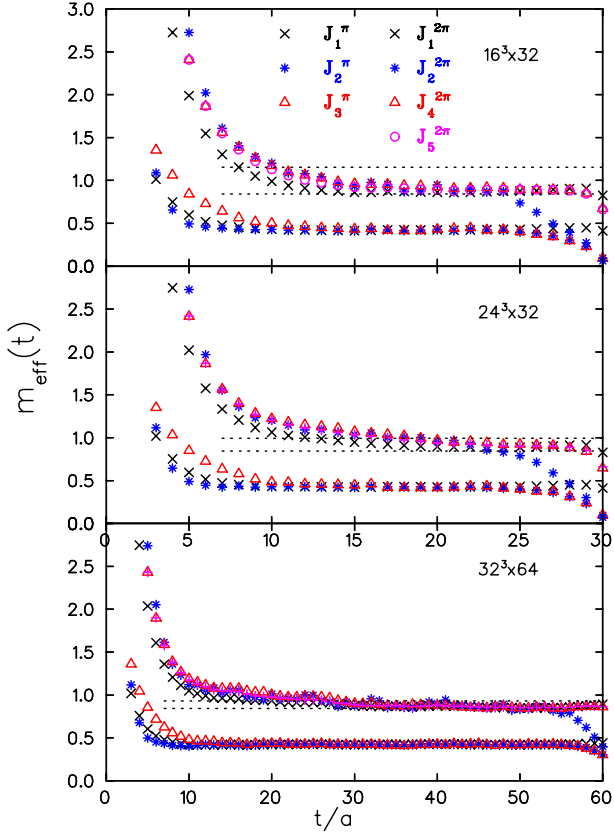}}
\caption{The effective mass at $\kappa_l=0.153$ as a function
of $t/a$ for the lattice of size $16^3\times 32$ (top),
$24^3\times 32$ (middle) and $32^3\times 64$ (bottom).
The set results with the lower value are obtained using one
pion interpolating
fields, ${\cal J}_{1}^\pi$ (crosses),   ${\cal J}_{2}^\pi$ (asterisks)
and   ${\cal J}_{3}^\pi$ (open triangles). The
set with the higher value correspond to the  two pion  
interpolating fields, ${\cal J}_{1}^{2\pi}$ (crosses), 
${\cal J}_{2}^{2\pi}$(asterisks), ${\cal J}_{4}^{2\pi}$ (open triangles), and 
 ${\cal J}_{5}^{2\pi}$ (circles). 
 The dotted lines are the two
lowest two-pion scattering states $E_{2\pi}^0$ and $E_{2\pi}^1$. }
\label{fig:mass two pions}
\end{figure}

Having tested our techniques for a single particle state we now verify that
they can be applied for two particle scattering states. For this test we choose
the two pion system in the isospin two channel where no low lying 
resonances are 
expected. We consider five interpolating fields:
\beq
{\cal J}_{1}^{2\pi}(x)&=& {\cal J}_1^\pi(x){\cal J}_1^\pi(x), 
\hspace*{0.5cm} {\cal J}_1^\pi(x)=\bar{d}(x)\gamma_5 u(x)  \nonumber \\
{\cal J}_{2}^{2\pi}(x)&=&  {\cal J}_2^\pi(x){\cal J}_2^\pi(x), 
\hspace*{0.5cm} {\cal J}_2^\pi= \bar{d}(x)\gamma_5 \gamma_0 u(x) \nonumber \\
 {\cal J}_{3}^{2\pi}(x)&=& {\cal J}_3^\pi(x){\cal J}_3^\pi(x), 
 \hspace*{0.5cm}{\cal J}_3^\pi=\bar{d}(x)\gamma_5 \hat{e}_\mu \gamma_\mu u(x) \nonumber \\
{\cal J}_{4}^{2\pi}(x)&=& {\cal J}_0^\rho(x){\cal J}_0^\rho(x), 
\hspace*{0.5cm}{\cal J}_0^\rho= \bar{d}(x)\gamma_0\sum_{i=1}^3\gamma_i u(x)\nonumber \\
{\cal J}_{5}^{2\pi}(x)&=& \sum_{i=1}^3  {\cal J}_i^\rho(x){\cal J}_i^\rho(x),
\hspace*{0.15cm}{\cal J}_i^\rho(x)=\bar{d}(x)\gamma_i u(x) 
\label{J two pions}
\eeq
The first three are products of pion interpolating fields 
 whereas the last two
are products of rho-type interpolating fields.
As we already pointed out the energy of a  scattering state of two
hadrons $h_1$ and $h_2$ depends
on the spatial size L of the lattice and is given by
\small
\be
E_{h_1h_2}^n=\sqrt{m_{h_1}^2+n \left(\frac{2\pi}{L}\right)^2} +
\sqrt{m_{h_2}^2+n \left(\frac{2\pi}{L}\right)^2}, \hspace*{0.1cm} n=0,1,\cdots
\label{energy spectrum}
\ee
\normalsize
where we have suppressed the $L$ index on the energy.
In Fig.~\ref{fig:mass two pions} we show the effective mass for a single pion 
using 
 the interpolating fields  ${\cal J}_1^\pi(x)$, ${\cal J}_2^\pi(x)$,
and ${\cal J}_3^\pi(x)$ for our three lattices. As expected 
${\cal J}_1^\pi(x)$, routinely used
in lattice calculations, and its variant  ${\cal J}_2^\pi(x)$
 have the largest overlap with the pion. All three yield consistent 
results for the pion mass. For the two pion system the interpolating field
${\cal J}_3^{2\pi}(x)$ is very noisy as compared to the other four
and we do not include it in the figure. We obtain the best overlap 
with the  two pion
ground state when using ${\cal J}_1^{2\pi}(x)$ with 
 ${\cal J}_4^{2\pi}(x)$ and  ${\cal J}_5^{2\pi}(x)$  being the next best.
Again all interpolating fields produce consistent two pion energies for large
enough  time separation. For lattices $16^3\times 32$ and $32^3\times 64$
the time extent is sufficient to have 
a large plateau region after suppression of excited state contributions.
For the $24^3\times 32$ lattice this occurs for time separations $t/a>22$
limiting the fit range to a few time slices.

\begin{figure}
\epsfxsize=8truecm
\epsfysize=11.truecm
\mbox{\epsfbox{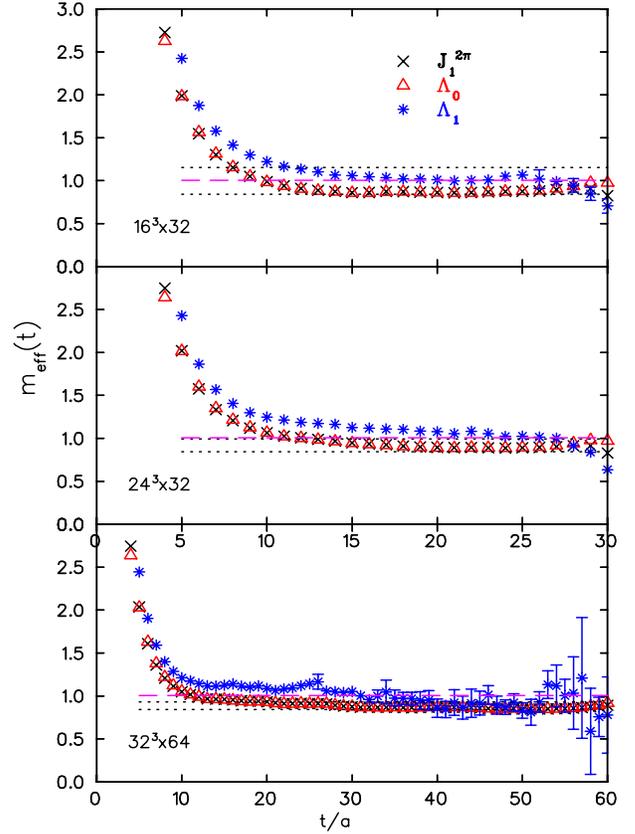}}
\caption{The effective mass at $\kappa_l=0.153$ extracted from the
correlator using
the  interpolating
field ${\cal J}_{1}^{2\pi}$ (crosses) and from the two largest
eigenvalues $\Lambda_0$ (open triangles) and $\Lambda_1$ (asterisks)
 for the lattice of size $16^3\times 32$ (upper graph),
$24^3\times 32$ (middle graph) and $32^3\times 64$ (lower graph). 
The dotted lines are the two
lowest energy two-pion scattering states $E_{2\pi}^0$ and $E_{2\pi}^1$. 
The dashed line is the lowest energy two-rho scattering state $E_{2\rho}^0$. }
\label{fig:eig two pions}
\end{figure}

\begin{figure}
\epsfxsize=8truecm
\epsfysize=6.truecm
\mbox{\epsfbox{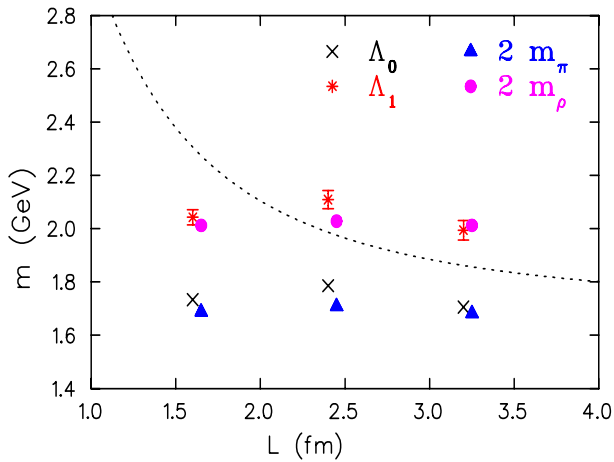}}
\caption{The mass at $\kappa_l=0.153$
extracted from  fitting the plateau of
the effective mass of the  lowest energy eigenvalue (filled triangles) and
 the second higher energy eigenvalue (filled circles shifted to right for 
clarity)
as a function of the spatial length of the lattice $L$.
Also shown are results
for twice the mass extracted from the pion correlator (crosses)
and rho correlator (asterisks). 
The dotted line is the two-pion scattering energy $E_{2\pi}^1$.}
\label{fig:eig vol}
\end{figure} 
\begin{figure}
\epsfxsize=8truecm
\epsfysize=11.truecm
\mbox{\epsfbox{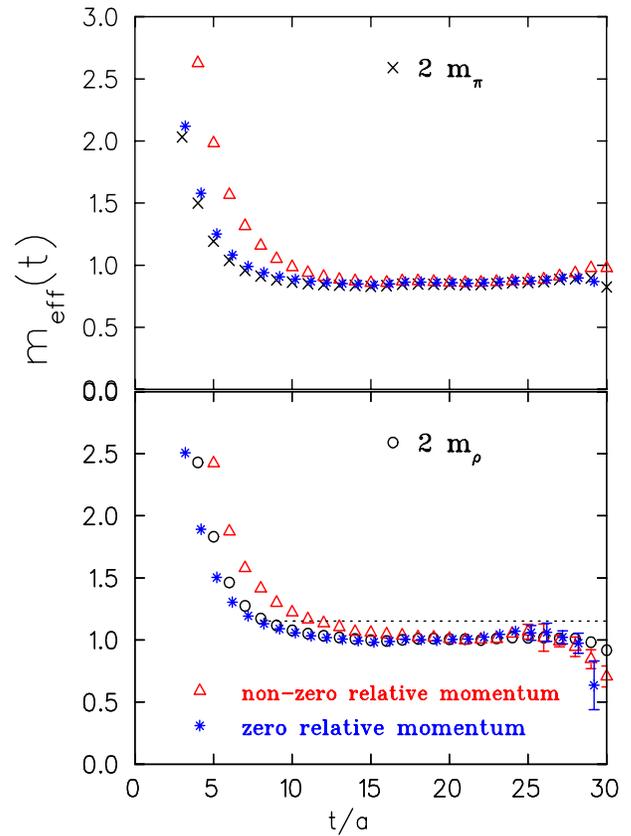}}
\caption{The upper graph shows the effective mass 
for the lowest energy
eigenstates and the lower graph the effective mass
for the second higher energy eigenstate. In both graph we show data
with projection to zero relative momentum using Eq.~\ref{p projection}
 (asterisks shifted to the right for clarity)
and without explicit projection (open triangles)
for the lattice of size $16^3\times 32$ at $\kappa_l=0.153$.
 We also show twice the effective mass extracted from 
a single pion (crosses in  upper graph ) 
and rho correlator (circles in lower graph).
The dotted line is the two-pion scattering state  $E_{2\pi}^1$. 
 }
\label{fig:eig project0}
\end{figure}

\begin{figure}
\epsfxsize=8truecm
\epsfysize=6.truecm
\mbox{\epsfbox{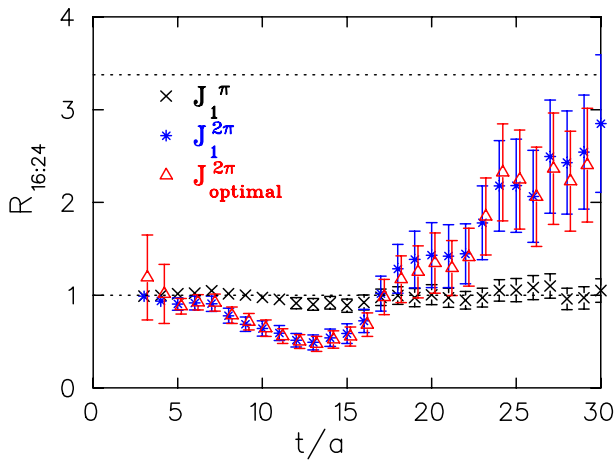}}
\caption{The ratio $R_{16:24}$  at $\kappa_l=0.153$
for one and two pion states using interpolating field ${\cal J}_1^\pi(x)$
(crosses) and
${\cal J}^{2\pi}_1(x)$ 
(asterisks) as a function
of $t/a$. The ratio $R_{16:24}$ extracted from correlators using
the optimal combination $J_{optimal}(x)$ 
for the lowest eigenstate of the two pion system is shown by 
the open triangles shifted to the right for clarity. 
The dotted lines show the expected value of the ratio for a 
single particle state and two-particle scattering state.}
\label{fig:corr ratio}
\end{figure}

\begin{figure}
\epsfxsize=8truecm
\epsfysize=11.truecm
\mbox{\epsfbox{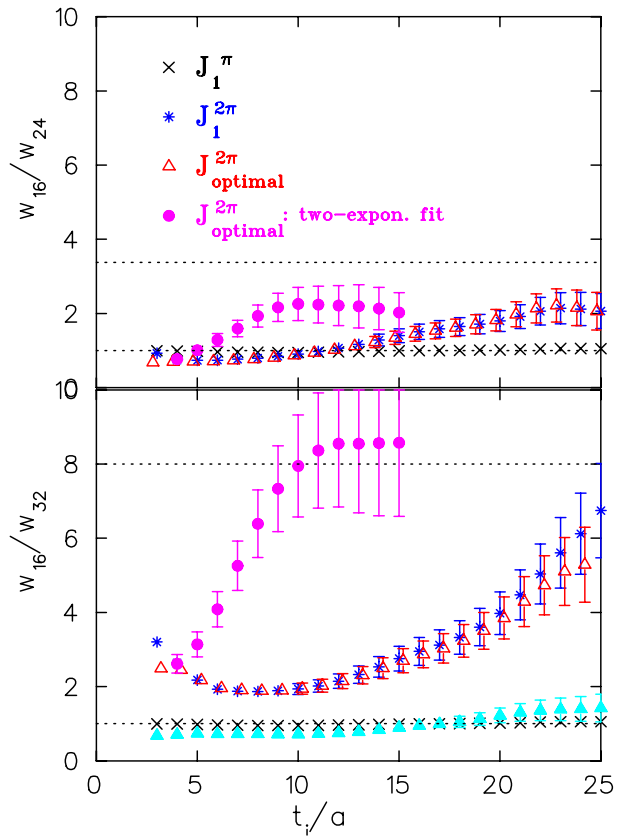}}
\caption{The ratio of spectral weights $w_{16}/w_{24}$ (upper graph)
and $w_{16}/w_{32}$ (lower graph) as a function
of the lower fit range $t_i/a$ at $\kappa_l=0.153$. Asterisks denote
results extracted from the correlators with ${\cal J}_1^{2\pi}(x)$,
open triangles and filled circles from correlators with
 $J_{optimal}(x)$ fitted to a single exponential 
to a sum of two exponentials respectively.
For comparison we also show these ratios for the single pion state
using ${\cal J}_1^\pi(x)$
(crosses).
The upper fit range is fixed
to 26 for the $16^3\times 32$ lattice and to 56 for the $32^3\times 64$ lattice.
On the lower graph we show with the filled triangles
the ratio of spectral weights $w_{16}/w_{32}$
when the upper fit range for the large lattice is fixed to 26.
 The dotted lines show the expected value of the ratios for a 
single particle state and a two-particle scattering state.}
\label{fig:zratio}
\end{figure}

\begin{figure}
\epsfxsize=8truecm
\epsfysize=6.truecm
\mbox{\epsfbox{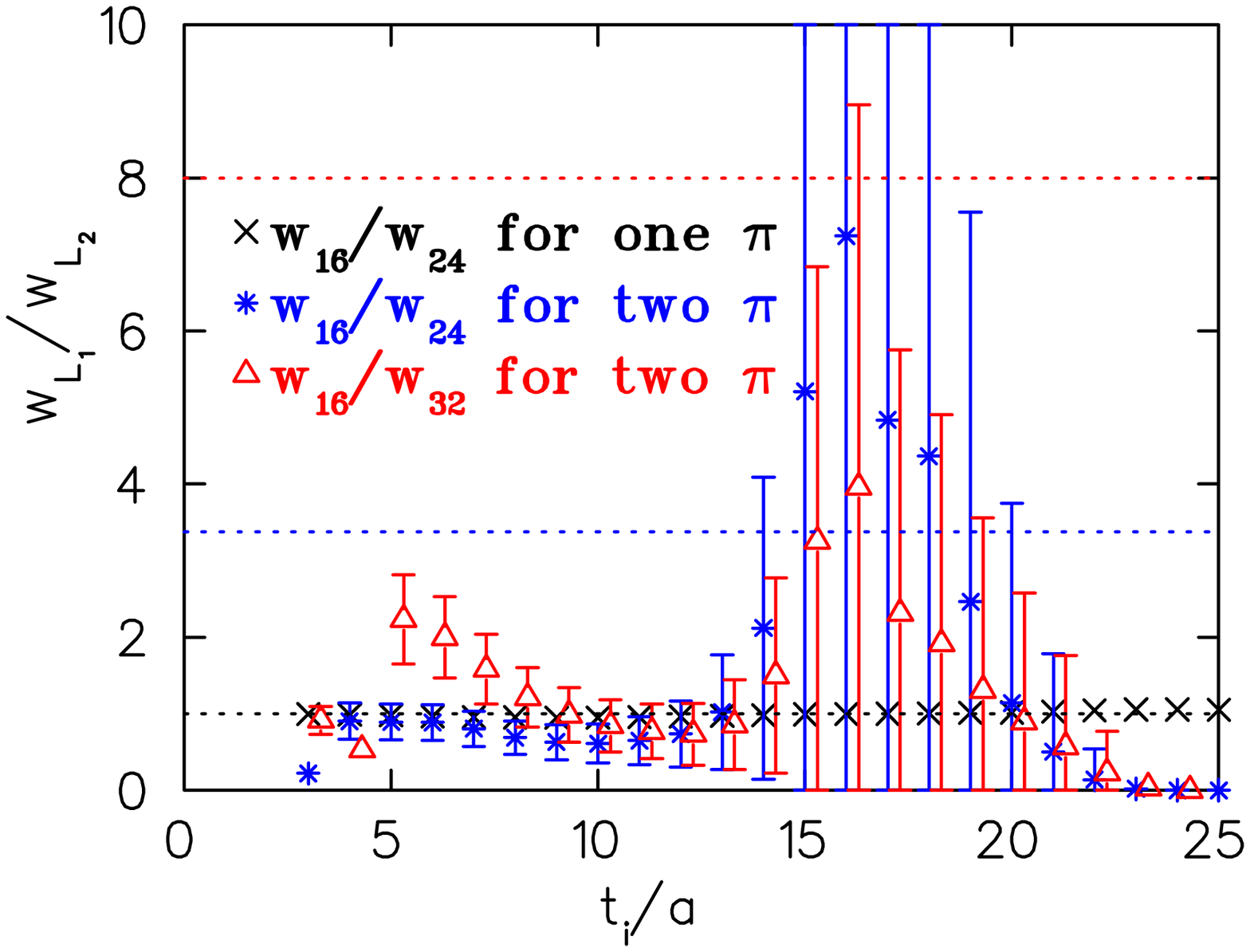}}
\caption{The ratio of spectral weights $w_{16}/w_{24}$
extracted  from the correlators with interpolating field 
 $J_{optimal}(x)$ (asterisks) for the
first excited state as a function of the lower fit range $i_i/a$.
The corresponding ratio $w_{16}/w_{32}$  is shown with the
open triangles. The dotted lines show the expected value of the ratio for a 
single particle state and two-particle scattering state.
The crosses show for comparison the ratio of spectral weights
for one pion state extracted from the correlator
 using interpolating field ${\cal J}_1^\pi(x)$.}
\label{fig:zratio1}
\end{figure} 

We perform a variational analysis using the interpolating fields
defined in Eq.~\ref{J two pions} excluding ${\cal J}_{3}^{2\pi}$, 
which means that in
Eq.~\ref{eigenvalue equation}  we take $M=4$. 
To check we also compute the eigenvalues
by the second method of analysis namely by projecting
using Eq.~\ref{project correlation} to an $N\times N$
matrix where we take $N=3$. What we find is that we obtain two distinct
 eigenstates for which we also construct the best interpolating
field using the eigenvectors ${\cal V}_n$.
For this system had we used only interpolating fields ${\cal J}_1$
and ${\cal J}_4$ the resulting eigenvalues 
would be consistent with
the ones obtained within the larger variational set.
This is an important observation since it demonstrates that using two
interpolating fields which have very good overlap with two low lying  
states we can accurately
determine them by diagonalizing the $2\times 2$ correlation matrix. 
 The effective
mass determined for the two eigenvalues $\Lambda_n$ is shown in
 Fig.~\ref{fig:eig two pions} for our three lattices.
In all cases the effective mass extracted from the lowest 
energy eigenvalue (largest
$\Lambda$)
is in agreement with that extracted from the correlator that
uses the interpolating field ${\cal J}^{2\pi}_1$. As the 
spatial extension of the lattice increases a larger time separation is needed
for the effective mass to reach the energy of the s-wave scattering
state $E^0_{2\pi}$. 
Since in this system
the low lying states are  two particle scattering states
this clearly demonstrates that the two lowest eigenvalues
do not correspond to a given value of the relative 
momentum carried by each particle but
they are an admixture of states having  non zero 
relative momentum that mix with the state with zero relative momentum. 
 The energy difference,  $E^n-E^{n-1}$, between different relative momentum 
states
decreases like $1/L^2$ and
therefore a larger time interval is needed in order to reach
the ground state as $L$ increases. This can  be seen 
in Fig.~\ref{fig:eig two pions} where for the $16^3$ lattice the plateau
for the ground state is reached for $t/a>13$ whereas for the $24^3$ we need
$t/a>23$ and for the $32^3$ $t/a>30$. These values determine the smallest
value of the lower time range to be used in fitting the effective
masses to a constant. 
If one would like to reduce the lower fit range then
 fitting to 
a two exponential form using Eq.~\ref{two exponential fit} is essential.
A similar behavior is also
observed for the second eigenstate: On the
$16^3$ the two-rho scattering state $E^0_{2\rho}$ is the second lowest
energy state since  $E^1_{2\pi}$ is higher and a clear plateau
is seen for $t/a>17$. Going to the larger $24^3$ lattice
 $E^1_{2\pi}$ decreases becoming about equal to  $E^0_{2\rho}$ 
with a plateau that sets in for $t/a\sim 25$ resulting in a 
fit range limited to three points. This explains
why the energy extracted is higher on this lattice
than $2m_\rho$ as can be seen in Fig.~\ref{fig:eig vol} where
we plot the mass extracted from fitting
the effective mass in the available
plateau region versus the spatial lattice size.
 Finally,  despite the fact that on the $32^3$ lattice
$ E^1_{2\pi}< E^0_{2\rho}$ the two-rho state
dominates the time dependence of the second eigenvalue  for $30\le t/a\le 40$.
For $t/a>40$ the effective mass becomes consistent
 with $E^1_{2\pi}$ but in this time range
  the statistical errors have become much larger.
 Therefore fitting the effective mass in a time range $30-40$ or
extending the upper fit range affects very little the resulting value
thereby obtaining the rather accurate measurement of
 $E^0_{2\rho}$ displayed  in Fig.~\ref{fig:eig vol}.
Note that corrections due to  
interactions between the two pions cannot be seen on the scale
 of this figure
assuming that we can use the physical scattering length in Luscher's 
result~\cite{Luscher scatter}. 
It must be noted that on the large lattice fitting in the
plateau-like region in the time range  $15<t/a<30$ would yield
an incorrect higher value for the energy. Therefore this demonstrates that
for lattices of spatial extent of about 3~fm we need a time separation of
at least $40a$ to obtain correctly the energy of the two-rho state making the
use of Dirichlet b.c. essential. 
To verify that the reason the plateau region starts at relative large time
separations 
 is due, to a large extent, to   contamination of states 
with higher relative momentum,
 we perform on the small lattice
an explicit momentum projection to two particles each carrying zero 
momentum. This is done by evaluating the correlation matrix 
\be
{\cal C}_{j^{s^\prime}k^{s}}(t)=\sum_{\bf x,y} 
<0|J_j^{s^\prime}({\bf x},t)J_j^{s^\prime}({\bf y},t)J_k^{s \dagger}(0) 
                                       J_k^{s \dagger}(0) |0> \quad,
\label{p projection} 
\ee
where $s,s^\prime$ denotes the pion or the rho.

In Fig.~\ref{fig:eig project0} we compare the eigenvalues obtained when we
carry out the zero momentum projection for each particle to
our previous (unprojected) results. As can be seen 
for  both  eigenstates the plateau region starts at earlier
time separations. In fact for $t/a>6$ the
projected two particle correlator is the same
as the product of the single particle correlators.
 For large enough time separations the unprojected eigenvalues
approach the correct values $E^0_{2\pi}$ and $E^0_{2\rho}$.
The mass gap $E^0_{2\rho}-E^0_{2\pi}\sim 320$~MeV
at  $\kappa_l=0.153$. The mass
gap between the $\Theta^+$ and the KN s-wave
scattering state is not known at this value of $\kappa_l$.
As will be discussed below, the mass gap between our
candidate pentaquark resonance and $E_{KN}^0$ increases with the quark
mass to about 170~MeV at $\kappa_l=0.153$ in the
negative parity channel. This  estimated smaller
gap 
can  make the study of the pentaquark system harder.

Having identified the two lowest eigenstates within our variational basis
we study the spectral weights of these states. 
In Fig.~\ref{fig:corr ratio} we show the ratio $R_{16:24}$ for the pion
 when using the interpolating field 
${\cal J}_1^\pi(x)$
and for two pions when using ${\cal J}_1^{2\pi}(x)$ as well
as  ${\cal J}_{optimal}$. This ratio is unity for the pion as
expected  whereas for the two-pion system
it increases approaching the expected ratio of $3.4$
for $t_i/a>25$ when the s-wave two pion scattering state
dominates. For the lattice of spatial extent $32$  this happens
for $t/a>30$ and therefore  $R_{16:32}$ 
stays close to unity up to about $t/a=30$ which is the largest time
separation for which it can be constructed.    
Instead 
of $R_{16:32}$ we can extract the  spectral weights by 
fitting the correlator to
a single exponential or to a sum of two  exponentials. 
This allows to take into account information
from the full time extent of the lattice. We fix the upper fit range to
26 in lattice units
 for the lattices of temporal extent $N_t=32$ and to 56 for the lattice
with $N_t=64$. 
We show in Fig.~\ref{fig:zratio}  
the ratio of spectral weights $W_{16}/W_{24}$ and $W_{16}/W_{32}$ 
for the lowest state.   
What can be seen is that both ratios 
extracted using a single exponential fit  increase
 approaching the expected value for a scattering state.
Fitting to a sum of two exponentials we obtain ratios that approach
the expected value at much smaller time separations albeit with larger
errors. 
 However if
instead of 56 we take an upper fit range of 26 for the large lattice
either using single or double exponential fits
the ratio $W_{16}/W_{32}$ stays close to unity leading to the wrong 
conclusion.
We show the corresponding ratios for the second eigenstate in 
Fig.~\ref{fig:zratio1} using
 a single exponential. Even though the errors on the correlators are small
the error on the ratio of spectral weights is too 
large to lead to a definite
conclusion. Had we used a sum of two exponentials the errors would be 
even larger.    
  
\subsection{The pentaquark sector}

The first experimental indication for the existence
of a pentaquark state came when the LEPS collaboration~\cite{LEPS}
detected a resonance with  
 mass $1540\pm 10$~MeV and width less than $25$~MeV in
accord with the predictions of the chiral soliton model~\cite{soliton}. 
This small width is surprising since, lying about 100 MeV above 
the KN threshold, it is expected to readily decay to KN. 
Since the observation of the first 
signal for $\Theta^+$ was reported, several other experiments were 
carried out, some confirming its existence~\cite{positive} 
and others not~\cite{negative,negative2}.

\begin{figure}
\epsfxsize=8truecm
\epsfysize=11.truecm
\mbox{\epsfbox{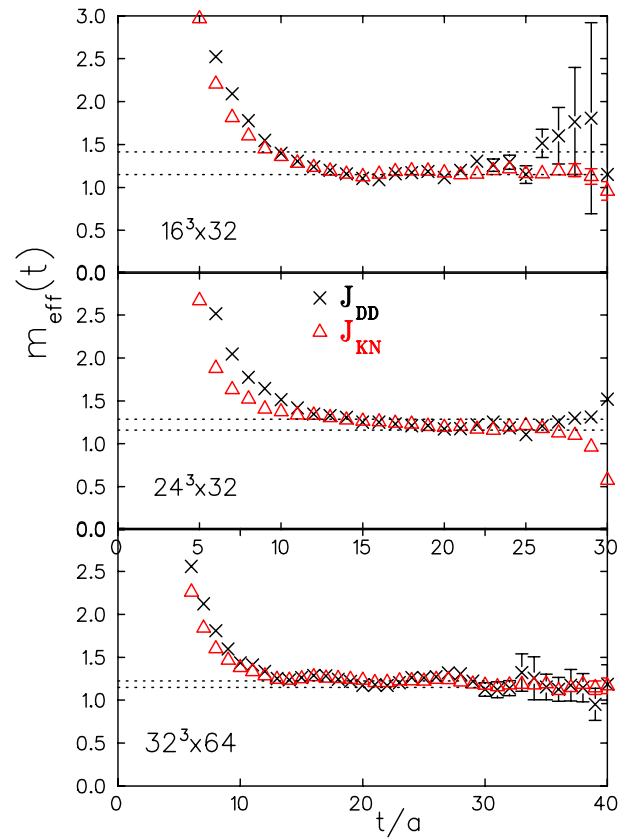}}
\caption{The effective mass for the pentaquark in the
negative parity channel
for lattices of size $16^3\times32$ (upper graph), $24^3\times 32$
(middle graph)  and $32^3\times 64$ (lower graph) at $\kappa_l=0.153$.
The crosses show the results obtained from correlators 
using ${\cal J}_{DD}$ and
the open triangles using  ${\cal J}_{KN}$. The 
 dotted
lines show $m_N+m_K$ and $E_{KN}^1$.}
\label{fig:theta_minus_meff}
\end{figure}

\begin{figure}
\epsfxsize=8truecm
\epsfysize=11.truecm
\mbox{\epsfbox{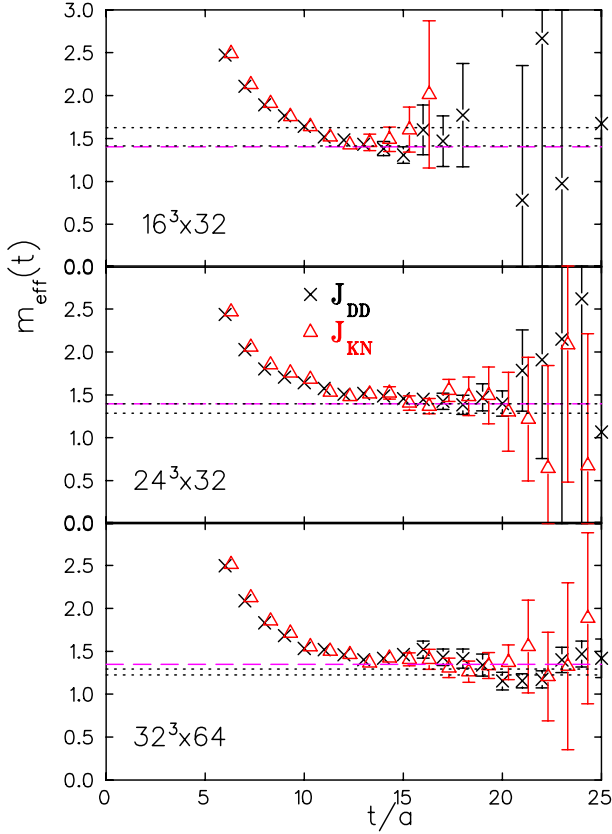}}
\caption{The effective mass for the pentaquark in the positive parity 
 channel at $\kappa_l=0.153$. 
The dotted lines show 
$E^1_{KN}$ and $E^2_{KN}$ and
the dashed line $m_K+m_{N^*}$.
The rest of the notation is the same as that of Fig.~\ref{fig:theta_minus_meff}.}
\label{fig:theta_plus_meff}
\end{figure}

\begin{figure}
\epsfxsize=8truecm
\epsfysize=11.truecm
\mbox{\epsfbox{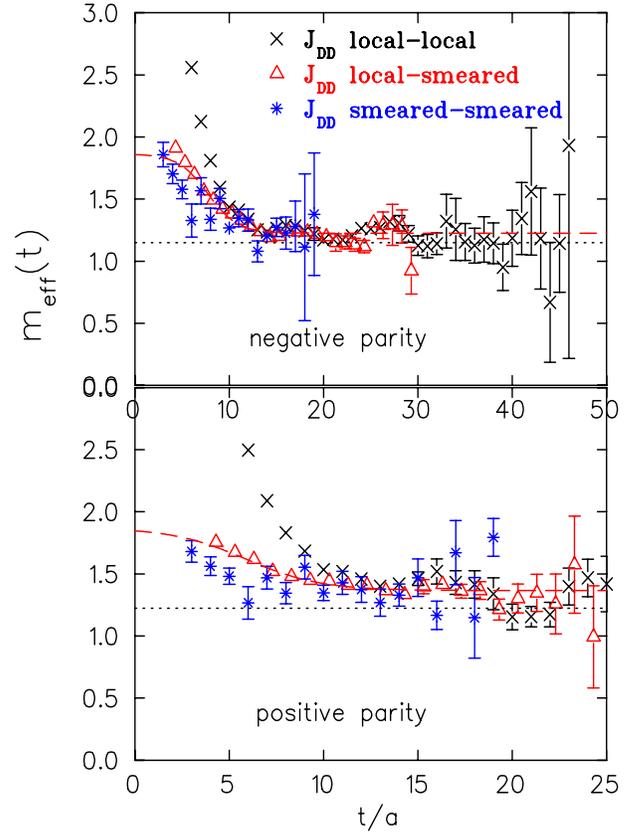}}
\caption{The upper graph shows the $\Theta^+$ effective mass 
for the negative  parity channel  and
the lower graph for the positive  parity
channel at $\kappa_l=0.153$ on the lattice of size
$32^3\times 64$. The crosses show results obtained 
from the local-local correlator $C(t)$, 
 the open triangles results from the 
local-smeared correlator $\tilde{C}(t)$ shifted to the right
and the asterisks results from
the smeared-smeared correlator  $\widehat{C}(t)$. The diquark-diquark
interpolating field
is used in all correlators. The
dashed line is a two-exponential fit to the local-smeared results. 
The dotted line
shows  $E_{KN}^0$ for the negative  and $E_{KN}^1$ for the
positive parity channel.}
\label{fig:theta_meff_32x64}
\end{figure}

\begin{figure}
\epsfxsize=8truecm
\epsfysize=9.7truecm
\mbox{\epsfbox{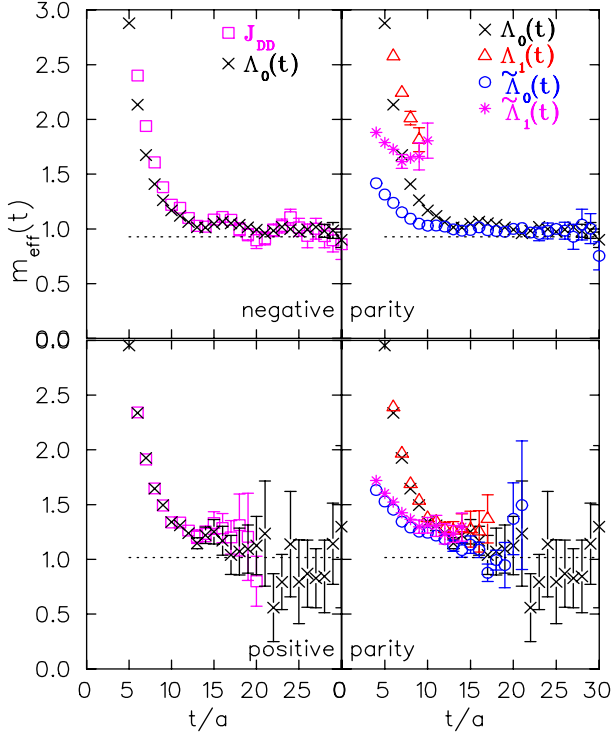}}
\caption{
The upper graphs show the  effective mass for the negative parity and
the lower graphs for the positive parity
channels at $\kappa_l=0.155$ on the lattice of size
$32^3\times 64$. 
The squares on the left two graphs denote the results
extracted from the correlators with 
 the local interpolating field ${\cal J}_{DD}$. The crosses 
on all four graphs show the effective mass, 
derived from the eigenvalue $\Lambda_0(t)$ of ${\cal C}_{DD;KN}$.
 On the graphs on the right the open 
triangles denote results derived from
$\Lambda_1(t)$, whereas  the circles and asterisks   are results 
extracted from the eigenvalue $\tilde{\Lambda}_0(t)$ and
  $\tilde{\Lambda}_1(t)$
  from the diagonalization of $\tilde{\cal C}_{DD;KN}$ respectively.
The dotted lines
show  $E_{KN}^0$ for the negative parity channel and $E_{KN}^1$ for the 
positive parity channel.}
\label{fig:correlation 0.155}
\end{figure}

\begin{figure}
\epsfxsize=8truecm
\epsfysize=11.truecm
\mbox{\epsfbox{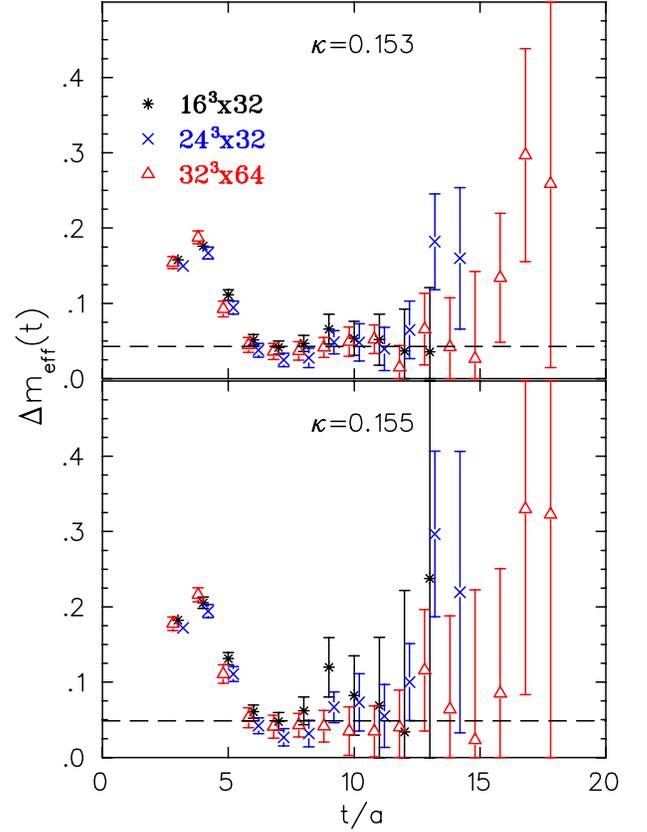}}
\caption{The energy gap between the two lowest energy eigenstates
as a function of $t/a$
  for the pentaquark in the positive parity
channel at $\kappa_l=0.153$ (upper graph)  and $\kappa_l=0.155$ (lower graph)
 for lattices of size $16^3\times 32$ (asterisks),  $24^3\times 32$ 
(crosses) and $32^3\times 64$ (open triangles). The dash line
shows the plateau value determined on the lattice of size $16^3\times 32$.} 
\label{fig:dmeff plus}
\end{figure} 

\begin{figure}
\epsfxsize=8truecm
\epsfysize=11.truecm
\mbox{\epsfbox{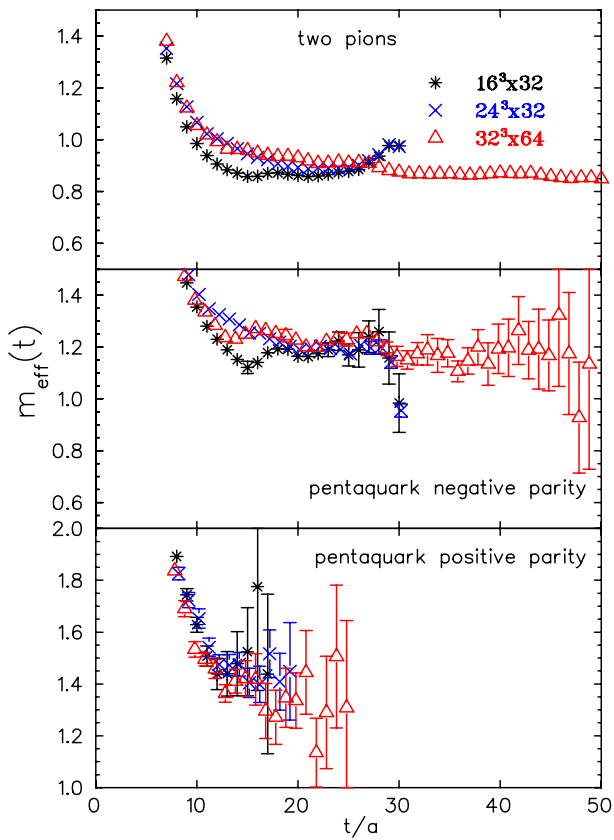}}
\caption{The upper graph shows the
  effective mass for the lowest energy eigenstate  
of the two-pion system as a function of $t/a$ at $\kappa_l=0.153$.
The middle graph shows the corresponding effective mass
 for the pentaquark system in the negative parity
channel  and the lower graph for the pentaquark 
in the positive parity channel 
 for the lattices of size $16^3\times 32$ (asterisks),  $24^3\times 32$ 
(crosses) and $32^3\times 64$ (open triangles).} 
\label{fig:meff}
\end{figure} 

\begin{figure}
\epsfxsize=8truecm
\epsfysize=6.truecm
\mbox{\epsfbox{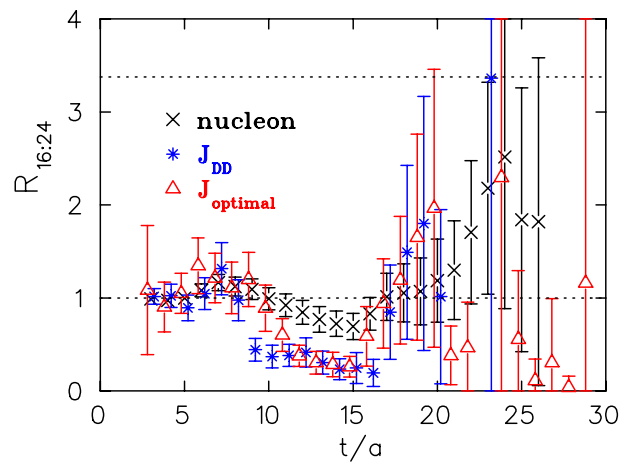}}
\caption{The  ratio 
$R_{16:24}$ versus $t/a$ for the negative parity channel
for the lowest energy eigenstate at $\kappa_l=0.153$. Asterisks show
results obtained
with  correlators using 
 the diquark-diquark interpolating
field  ${\cal J}_{DD}$ and open triangles with 
${\cal J}_{optimal}$. 
 For comparison we also show the
ratio for the nucleon (crosses).  
 The dotted lines show the expected value of the ratios for a 
single particle state and a two-particle scattering state.}
\label{fig:theta_zratio}
\end{figure}
 
\begin{figure}
\epsfxsize=8truecm
\epsfysize=11.truecm
\mbox{\epsfbox{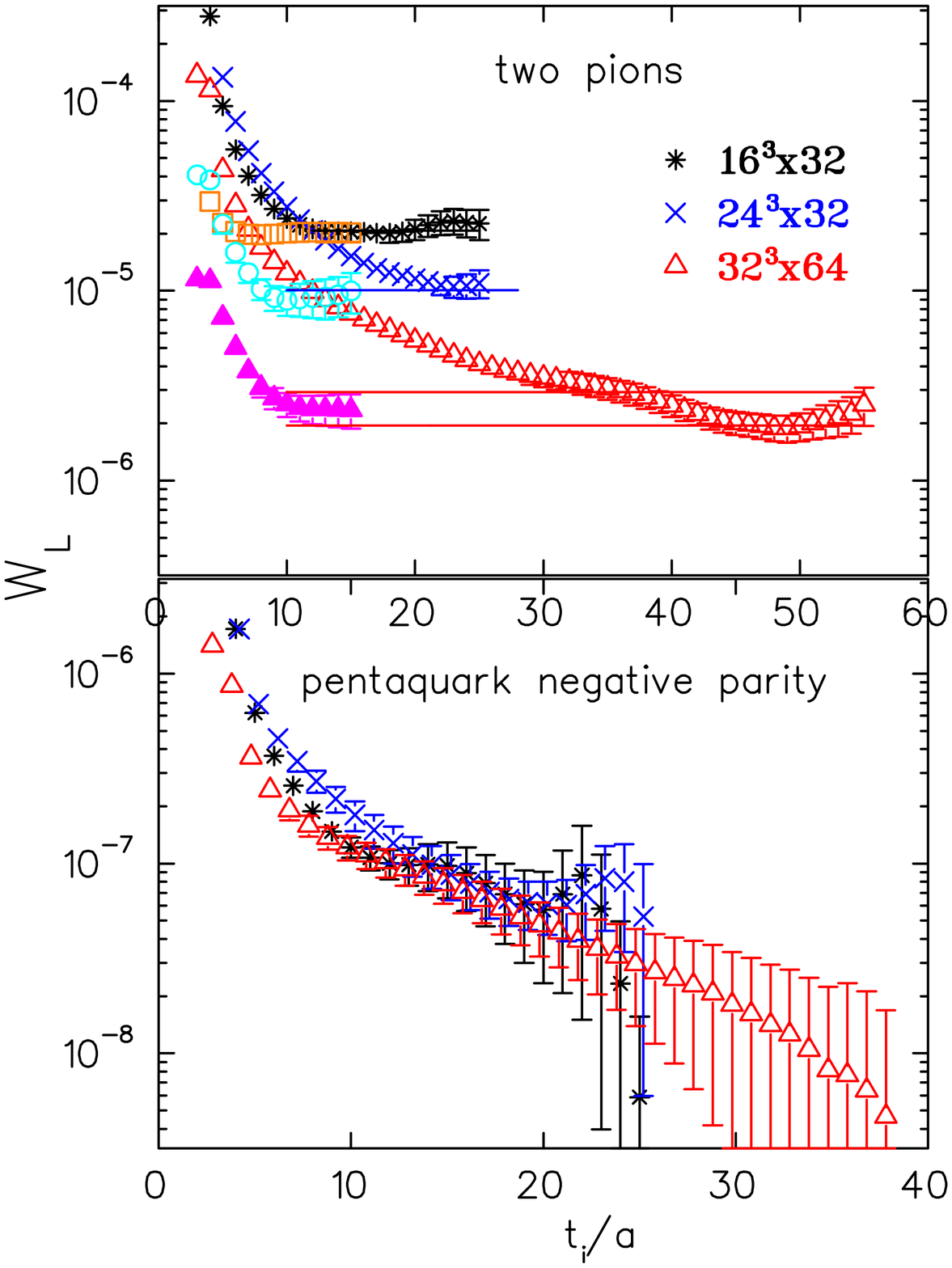}}
\caption{We show the spectral weights
 for the lowest eigenstate at $\kappa_l=0.153$ versus the lower 
fitting range $t_i/a$ keeping the upper range fixed to 26 for the
two smaller lattices and to 56 for the large lattice. 
The upper graph shows results for the two-pion system and the lower graph 
for the pentaquark in the negative parity
channel  
 for the lattices of size $16^3\times 32$
 (asterisks for a single
exponential fit), 
 $24^3\times 32$ 
(crosses for a single
exponential fit) 
and $32^3\times 64$ (open triangles for a single
exponential fit). In the case of the two pion system
we also show results for the weights determined from fitting the correlator
to a sum of two exponentials for lattices of size  $16^3\times 32$
 (squares),
 $24^3\times 32$ 
(circles) 
and $32^3\times 64$ (filled triangles). The solid lines in the upper
graph show the plateau values.  } 
\label{fig:zbest}
\end{figure}

Whether lattice QCD supports the existence of 
a pentaquark state is also unclear~\cite{summary}.
The study of resonances in lattice QCD is in its infancy and therefore
it comes with no surprise that identifying the $\Theta^+$ on the lattice 
turned out to be a difficult task.  In the two pion system
 scaling of the spectral weights $W$ with the
spatial volume was observed for large time separations. We
will use the largest available time separations in applying the same analysis
in the pentaquark system. Clearly whether this criterion  can be used
in practice 
will depend on the size of the statistical errors.

In Figs.~\ref{fig:theta_minus_meff} and \ref{fig:theta_plus_meff}
we show the effective mass for both the negative and positive parity 
states evaluated on the three lattices at $\kappa_l=0.153$ for
isospin zero using
the local interpolating fields ${\cal J}_{DD}$ and ${\cal J}_{KN}$.
What can be seen from  these figures
is that both interpolating fields yield consistent plateaus in both channels
with the KN interpolating field yielding  smaller errors in
the negative parity channel than
 the diquark-diquark interpolating field whereas in the positive parity channel
the opposite is true.  
We also see that the errors even at this
heavy pion mass are large especially for the positive parity channel
where the large time behavior is not well determined. This means
that extracting  the spectral weights for the positive
parity channel accurately enough for large time
separations to test scaling will not be possible. Therefore
for the positive parity channel we can only look for   volume dependence
in the extracted energy of the states. 
This energy can be determined
 either by fitting the effective mass in the plateau
range to a constant  or,
allowing for two states,  
to the form given in Eq.~\ref{two exponential fit}. 
The latter  is especially important when using local-local
 correlators, where 
excited states have a large contribution
up to  time separation of  about  $14a $. Since the aim
is to detect a resonance, which is higher than the scattering KN
state, it is crucial to  be able to fit in a time range where it is still the 
dominant state. This is expected to be so for time intervals less than 
about $t/a\sim 20$ or  $t\stackrel{<}{\sim} 10$~GeV$^{-1}$.
The values 
obtained for the mass, fitting within this range
 using  either ${\cal J}_{DD}$ or ${\cal J}_{KN}$,
are given in Table~\ref{table:mass 16_24} for all three lattices.
From these values, we can conclude that  either
 using
the diquark-diquark or the KN local interpolating field  the masses 
that we find are  within each other's error.
The systematic volume effect of a 
decrease in the mass with  the increase of the lattice size,
 characteristic of a scattering state,
is not observed in the positive parity channel.

The standard approach
to suppress excited state contributions improving the effective mass
  plateaus  
is to use smearing. 
We use Wuppertal smeared interpolating fields  on the large lattice 
 imposing antiperiodic boundary conditions in the temporal
direction.
As can be seen in Fig.~\ref{fig:theta_meff_32x64},
where we show results for
 $\kappa_l=0.153$, smearing the source but keeping the sink local 
suppresses excited state contributions. Note that,
within our statistical errors, the effective mass with smeared source on
half the lattice is in agreement with the results
from the local correlator at larger times  leading to the same plateau
values. This justifies the use
of antiperiodic b.c. As we already pointed out 
if the sink is also smeared the gauge noise increases making
identification of two 
close-by states impossible. Therefore for the 
rest of the results we will apply smearing only at the source.

Having confirmed that both local and smeared interpolating fields as well
as the KN and diquark-diquark interpolating fields lead to consistent values
for the mass we perform the variational analysis described in section II.
As in the two-pion system the eigenvalues will
still be contaminated by scattering states with non-zero relative momentum.
 This is particularly 
severe on the large lattice where the scattering KN states are very 
close in energy as can be seen
in Figs.~\ref{fig:theta_minus_meff} and \ref{fig:theta_plus_meff}. 
If within our variational basis we have interpolating
fields that have good overlap with the pentaquark resonance then
the two energy eigenvalues should yield the KN scattering states and
the $\Theta^+$. The choice of   the diquark-diquark interpolating
field  is based on the expectation that it should approximate well the
structure of $\Theta^+$. However if this is not the case then
we will not be able to obtain the resonance state as the second eigenvalue
of the $2\times2$ correlation matrix. The lowest 
eigenstate will be accurately determined  
and, allowing for  
large enough time separations, it should
produce the s-wave KN scattering state in the negative channel and
the p-wave KN scattering state in the positive channel.
If there is an admixture of a resonance state 
 then we expect the spectral weights to   show
a different behavior from that observed in the two-pion system
where the only low lying states are scattering states. 
In the positive parity channel
only KN scattering states with non-zero relative momentum contribute.
Therefore the lowest energy is expected to decrease
 as the spatial volume of the lattice increases
according to Eq.~\ref{energy spectrum}.
 Since we expect the errors on 
 the spectral weights to be too large in this channel
to allow us to test scaling
the only option is to study the volume
dependence of the energy  on the spatial
length of the lattice. 
However  in the positive parity channel
the s-wave $KN^*$ scattering state is  allowed. The energy of this state is 
 shown in Fig.~\ref{fig:theta_plus_meff}. 
On the small lattice the value of the smallest relative
momentum $2\pi/L$ is such that, at $\kappa_l=0.153$ using
the masses given in Table~\ref{table:parameters}, $E^0_{KN^*}$
 comes out  very close to $E_{KN}^1$ 
whereas  for the two larger lattices
is higher. 
Therefore if the two lowest eigenstates are the $KN^*$ and $KN$
scattering states  then we expect the
energy gap $|E^1_{KN}-E^0_{KN^*}|$ while being almost  zero on the 
small lattice
 to increase for the other two. 
Clearly   
 the existence of the $KN^*$ scattering state complicates
the identification of a pentaquark resonance also in the positive parity
channel and would require at least a  $3\times 3$ or even a $4\times 4$
correlation matrix to allow us to resolve three different
states. A very accurate determination of the spectral weights
for the three lowest states will be needed in order to distinguish
 the KN and KN$^*$ scattering states from  a resonance. This is 
beyond the scope of this work. 
 Our working hypothesis is that if the  $\Theta^+$ is present 
then it should dominate the correlator in the appropriate time range 
and to have largest coupling to the KN scattering state.

In Fig.~\ref{fig:correlation 0.155}  
we  show effective masses
  determined from the two  eigenvalues $\Lambda_0(t)$ and $\Lambda_1(t)$ 
of the correlation matrix $C(t)_{DD;KN}$ for  
$\kappa_l=0.155$ corresponding to a lighter quark mass than
 $\kappa_l=0.153$ discussed up to now  although
the  behavior is similar.  
We observe that, like in the case of the two-pion system, the effective
mass derived from the lowest energy eigenvalue is consistent with   that
obtained from either interpolating field ${\cal J}_{DD}$ or ${\cal J}_{KN}$.
On the same figure we
also show the effective mass obtained from the eigenvalues 
$\tilde{\Lambda}_0(t)$ and $\tilde{\Lambda}_1(t)$ 
of the correlation matrix $\tilde{C}(t)_{DD;KN}$, which involves
local-smeared correlators of the two interpolating fields 
 ${\cal J}_{DD}$ and  ${\cal J}_{KN}$  and antiperiodic boundary
conditions in the temporal direction.
 Again both local-local 
and local-smeared results yield the same plateaus with smearing
suppressing excited state contributions.
The main observation is that
the first excited state behaves differently in the positive and negative
parity channels:
 In the negative parity channel is very high in energy and  poorly
determined. This is 
puzzling since if the diquark-diquark operator has good overlap with
the $\Theta^+$ then we
would have expected 
it to be the first excited state of this correlation
matrix with the ground state being the KN scattering state.
In fact if we compare
 the overlap of ${\cal J}_{DD}$ with the ground state
to that obtained with ${\cal J}_{KN}$ we find
that it is about 50 times smaller.
This raises questions as to how
well  a local diquark description approximates the structure of
a pentaquark resonance. 
 In the positive parity channel we obtain
two distinct eigenstates very close together. From our
study of the two pion system we know that the eigenvalues,
in general, will not have a definite relative momentum but
will be contaminated with states with higher relative momenta. Therefore we
do not expect that for the time
separations that we can determine the gap to be able to  
relate it to $E^{2}_{KN}-E^{1}_{KN}$.
What we can do is to
check if this gap depends on the spatial size of the lattice.
We 
evaluate this energy gap on our three lattices at two values
of $\kappa_l=0.153$ and $\kappa_l=0.155$ and plot the results  
in Fig.~\ref{fig:dmeff plus}. Within our statistical accuracy
 the energy gap does not show any strong volume dependence  
 having about the same values of  100~MeV at the two $\kappa$ 
values. 
Although this energy gap agrees with the energy difference between
the $\Theta^+$ and the KN scattering threshold the mass 
that we  estimate below  for the positive channel at the chiral limit
is much larger than 1540 MeV to be identified as the $\Theta^+$.

Fitting the effective
mass derived from  $\Lambda_0(t)$
to Eq.~\ref{two exponential fit}
we find at $\kappa_l=0.153$
$E_0^{-}=1.228(14)$ for the negative parity
and $E_0^{+}=1.438(60)$ for the 
positive parity channels whereas    at $\kappa_l=0.155$ we find 
$E_0^{-}=1.014(20)$ and $E_0^{+}=1.259(116)$.
These values  agree both with those extracted from
the local-smeared correlator $\tilde{C}(t)$ using   ${\cal J}_{DD}$
given in Table~\ref{table:mass_smeared} and 
 from  $\tilde{\Lambda}_0(t)$ as demonstrated in 
Fig.~\ref{fig:correlation 0.155}. 
The fact that these 
different ways of  constructing the  correlation matrices
lead to the same value for the energy 
 makes us confident that our results 
for the lowest state 
are robust. For the second state the only 
estimate that we can provide is that, at
our two heaviest pion masses,  is higher in energy by
 about $ 100$~MeV 
in the positive parity channel.

In Fig.~\ref{fig:meff} we compare the effective masses for the lowest state
 of the pentaquark system and of 
the two-pion system. 
 The effective masses in the two-pion system start off
with higher values as the spatial volume increases because
scattering states
with higher relative momentum have smaller energy gap 
at larger spatial volumes, requiring
longer time  to damp out. Only at large times  the
ground state dominates the correlator yielding  the same plateau
value on all lattice sizes. 
 In the pentaquark system,  on the other hand, the
effective masses have the same value in the time interval $20<t/a<30$.
This is the time interval where a resonance 
is expected to dominate the correlator. 
However one has to bear in mind that the statistical
errors are bigger than in the two-pion system
 so that any volume dependence maybe  harder to detect.
This is particularly true in the positive parity channel. 
Therefore we 
  check if
 scaling of the spectral weights can yield information on
the nature of these states. For the same 
reasons as explained for the two-pion system we only show
the ratio of correlators $R_{16:24}$. As can be seen in 
Fig.~\ref{fig:theta_zratio}  for the negative
parity channel   $R_{16:24}$  stays
close to unity up to time separations $t/a\sim 20$.
For $t/a>20$ where, from our study of the two-pion system,  we expect 
this ratio to start to deviate from unity the errors become large making it
difficult to distinguish a single particle state
from a scattering state. In the positive channel in this large time
window
the errors are even larger making this test not applicable.
Instead of the ratio $R_{16:24}$ we plot the spectral weights 
determined on each volume
making a direct comparison to the two-pion system. Again we do this
only for the negative parity channel where we have more accurate results.
This comparison is shown in Fig.~\ref{fig:zbest} as a
function of the lower time range $t_i/a$ used in the single and
double  exponential 
fits
to the optimal correlators keeping the upper fit range fixed
to $26$ for the two smaller lattices and to $56$ for the 
large lattice.  For the two-pion system 
the weights are clearly volume dependent. 
On the lattice of size
$16^3\times 32$ the values we obtain, performing a single
exponential fit, are independent of the
lower fit range for $t_i/a>10$ and we can reliably extract 
the spectral weight of the lowest state. For 
the  $24^3\times 32$ lattice the  convergence is seen later because suppression
of higher momentum states requires larger time separations. For the 
$32^3\times 64$ lattice the 
 spectral weights are less well determined and
can vary by a factor of about two depending which range of $t_i/a$
we use. However, within this uncertainty, the spectral weight ratio
is closer to the values expected for a scattering state than to unity.
Fitting to a sum of two exponentials the spectral weights become independent
of the lower fit range for $t_i/a>10$ on all three lattices yielding
results consistent to those obtained when fitting to a single exponential.
This checks that indeed 
the value we find for the spectral weights does not depend
on our fitting scheme.
The volume dependence of spectral weights 
for the pentaquark system is different. 
The spectral weights extracted
from a fit to a single exponential~\footnote{ Fitting to a sum of two
exponentials yields spectral weights that are consistent with
those extracted from the single exponential fit but with errors that are too 
large to plot in Fig.~\ref{fig:zbest}. }
are volume independent for $10<t_i/a<26$, which is the
time range where we can make a comparison. This means
that in this range we do not have a single scattering state.
The energy gap between the two lowest KN scattering
states for $\kappa_l=0.153$ 
is $a(E_{KN}^1-E_{KN}^0)  = 0.26$ and $0.13$ on the lattices
of size $16^3\times 32$ and $24^3\times 32$ respectively as compared to  
 the energy gap of $a(2m_\rho-2m_\pi)=0.16$ in the two-pion system.
This means  
 that, as compared to the two-pion system, the lowest KN scattering
state is filtered out at smaller time separations on the smallest lattice
and at $\sim 20\%$    larger  times on the $24^4\times 32$ lattice. Therefore
one would have expected to see a clear
 volume dependence of the spectral weights
on the two smaller lattices which is not observed.  
On the largest lattice the fact
that the value of spectral weight
decreases as
 $t_i/a$ increases beyond 30 
is consistent with the expectation that at large time separations
a KN scattering state should dominate. 
 For the positive parity channel extending the upper
fit range to large time separations yields
 statistical errors that are too large  to make this test  
conclusive even at this very heavy 
pion mass. As the light quark mass
decreases the errors on the determination of the spectral weights 
increase and therefore one
would need much larger statistics to reliably extract the weights. Based on the
volume study  of the spectral weights at $\kappa_l=0.153$   we can
not exclude  a pentaquark resonance.

Within our variational basis the lowest energy eigenvalue is the
only one that we can determine accurately  
in the  negative parity channel. It is shown to be in  agreement with the 
mass extracted from the correlators using either local interpolating
fields  ${\cal J}_{DD}$,
${\cal J}_{KN}$ and ${\cal J}_{optimal}$ or the smeared versions
of these.
 Therefore to get an estimate on the light quark mass dependence 
of the lowest state we 
 use the local-smeared correlators $\tilde{C}(t)$.
 Similarly
for the positive parity channel the 
energy gap between the two lowest energy eigenvalues is 
about $100$ MeV at both $\kappa=0.153$ and $\kappa=0.155$.
Resolving these two states for smaller quark masses is even harder and
therefore we opt to evaluate just the lowest energy which
can again be obtained from just using $\tilde{\cal J}_{DD}$. This will
give us a rough idea of what the chiral limit of this state is.
Since the local-smeared correlators are evaluated on our largest volume
with periodic b.c. in the time direction the mass is extracted 
 using the form given in Eq.~\ref{two exponential fit} for which 
half the time separation is sufficient as demonstrated by
the agreement of our results at $\kappa_l=0.153$ and $0.155$.

\begin{widetext}

\begin{table}[ht]
\caption{The first column gives the
interpolating field, 
the second  column  gives the mass of the  pentaquark extracted by fitting 
the effective mass to a constant in the plateau region and the third  column  gives the mass
extracted by fitting to  
the form given in Eq.~\ref{two exponential fit} for the lattice of size $16^3 \times 32$. 
The corresponding quantities for the other two lattices are given in the next four columns.
We give results at values of $\kappa_l=0.153$ and $0.155$ 
for both parity states.
The $\chi^2/d.o.f$ for these fits is less than one.}
\label{table:mass 16_24}
\begin{tabular}{|c||c|c||c|c||c|c|} 
\hline
\multicolumn{1}{|c||} {} & 
\multicolumn{2}{|c||} {$16^3\times 32$} & 
\multicolumn{2}{|c||} {$24^3\times 32$} & 
\multicolumn{2}{|c||} {$32^3\times 64$} \\ 

\hline
\hline
\multicolumn{1}{|c||} {}&
\multicolumn{6}{|c|} {$\kappa_{l}=0.153$}\\
\hline
\multicolumn{1}{|c||}{} &
\multicolumn{6}{|c|} {Negative Parity} \\
\hline
 operator & 
 $am$ (1 exp)  & $am$ (2 exp) &  $am$ (1 exp) &  $am$ (2 exp) &
 $am$ (1 exp)  & $am$ (2 exp) \\
\hline
${\cal J}_{DD}$ 
& 1.172(14) & 1.174(14)   
& 1.236(14) & 1.240(15) &  1.237(12) & 1.234(13) \\   
\hline
 ${\cal J}_{KN}$ 
&  1.173(9)  & 1.180(9)    
&  1.232(15) & 1.239(17) &  1.239(9) & 1.236(9) \\
\hline
${\cal J}_{optimal}$
& 1.178(14)  & 1.176(16)
&  1.232(14)& 1.253(23)  &  1.235(9) & 1.230(9) \\
\hline
\multicolumn{1}{|c||} {}&
\multicolumn{6}{|c|} {Positive Parity} \\
\hline
 ${\cal J}_{DD}$ 
& 1.432(50) & 1.376(59)   
& 1.485(30) & 1.449(48) &  1.382(30) & 1.423(25) \\
\hline
 ${\cal J}_{KN}$ 
& 1.488(51) &  1.411(87)   
& 1.479(44) &  1.434(64) & 1.400(36) &  1.414(35) \\
\hline
${\cal J}_{optimal}$
& 1.502(135)  & - 
& 1.448(52) & 1.465(167) & 1.462(46) & 1.444(68) \\
\hline
\hline
\multicolumn{1}{|c||} {}&
\multicolumn{6}{|c|} {$\kappa_{l}=0.155$}\\
\hline
\multicolumn{1}{|c||} {}&
\multicolumn{6}{|c|} {Negative Parity} \\
\hline
 operator &
 $am$ (1 exp)  & $am$ (2 exp) &  $am$ (1 exp) &  $am$ (2 exp) &
 $am$ (1 exp)  & $am$ (2 exp) \\
\hline
${\cal J}_{DD}$ 
& 0.977(44) & 0.938(23) 
& 1.014(19) & 1.013(23) & 1.031(15) & 1.034(15) \\
\hline
 ${\cal J}_{KN}$ 
&  0.947(15)&  0.937(17) 
&  0.998(20)&  1.002(26) &  1.023(10) &  1.020(11) \\
\hline
${\cal J}_{optimal}$
& 0.952(19) & 0.929(27)
& 1.006(15) & 1.012(32) & 1.022(13) & 1.020(14) \\
\hline
\multicolumn{1}{|c||} {} &
\multicolumn{6}{|c|}{Positive Parity} \\
\hline
${\cal J}_{DD}$ 
& 1.300(67) & 1.184(92) 
& 1.323(33) & 1.316(44) & 1.308(26) & 1.250(49) \\
\hline
 ${\cal J}_{KN}$ 
& 1.374(60) & 1.283(129) 
& 1.274(48) & 1.255(67)  & 1.230(42) & 1.218(47) \\
\hline
${\cal J}_{optimal}$
& 1.457(89)  & - 
& 1.256(57) & 1.283(179) & 1.290(75) & 1.259(116) \\
\hline
\end{tabular}
\end{table}

\end{widetext}

\begin{table}[ht]
\caption{We give, in lattice units,  the mass of the pentaquark in the negative
and positive parity channels,
$m^-$
and $m^+$, 
determined  using Eq.~\ref{two exponential fit} to fit the 
effective mass extracted from the local-smeared correlator $\tilde{C}(t)$
with interpolating field ${\cal J}_{DD}$
on the lattice of size  $32^3\times 64$.
The fit range is  $5-20$ in lattice units and the values at the chiral
limit are obtained by linear extrapolation. It has been verified
that the values for the masses do not change outside 
errorbars if the lower fit range is increased from 5 to 7-10.}
\label{table:mass_smeared}
\begin{tabular}{|c|c||c|} 

\hline
\multicolumn{2}{|c||} { Negative Parity} & 
\multicolumn{1}{|c|} {Positive Parity} \\ 
\hline
 $\kappa_{l}$ & 
 $am^{-}$ &  $am^{+}$ \\
\hline
$0.153 $  
& 1.226(17) 
& 1.366 (35)\\
\hline
$0.1550 $  
&  0.999(21)
&  1.191(45) \\
\hline
$0.1554 $  
&   0.946(23)
&   1.167(48)  \\
\hline
$0.1558 $  
&  0.891(25) 
& 1.149(54) \\
\hline
$0.1562 $  
&  0.834(29)
&  1.135(63) \\
\hline
$\kappa_c $  
& 0.701(29)  
& 1.036(51)  \\
\hline

\end{tabular}
\end{table}

\begin{figure}
\epsfxsize=8truecm
\epsfysize=7.truecm
\mbox{\epsfbox{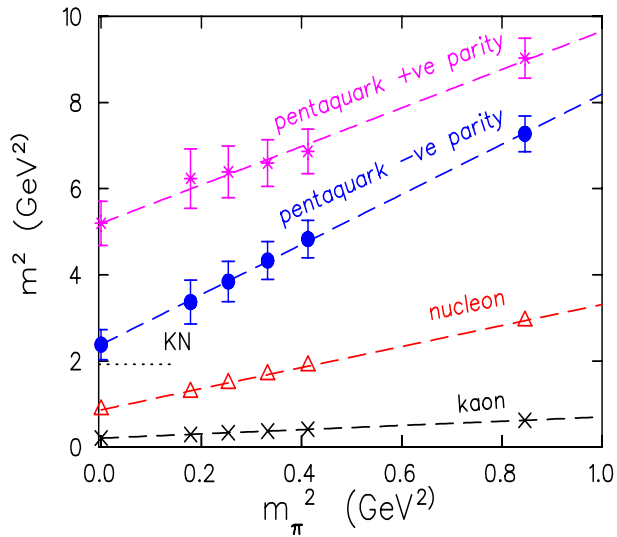}}
\caption{The mass squared of the kaon (crosses), nucleon (open triangles), 
 pentaquark in the negative parity channel (filled circles) and in the 
positive parity channel(asterisks)
 plotted
versus the pion mass squared. 
 We use the nucleon mass at the chiral limit to set the scale obtaining
 $a^{-1}=2.2$~GeV. 
The dashed lines show the
linear extrapolations to the chiral limit. The dotted line shows the KN 
threshold at the chiral limit.}
\label{fig:chiral32}
\end{figure}

The results that we obtain for the mass from the local-smeared
correlators
for all the light quark masses 
that we have considered in this work are given in 
Table~\ref{table:mass_smeared}. 
We note that
at $\kappa_l=0.155$ 
diagonalization of the correlation matrix $\tilde{\cal C}_{DD;KN}(t)$
yields in the negative parity channel $E_0^{-}=1.002(15)$ for the ground state 
and $E_1^{-}=1.6(2)$ for
the first excited state.
 This value of $E_0^-$ is  in  agreement with the value 
given in Table~\ref{table:mass_smeared} and  the value of $E_1^{-}$ is
in agreement with $1.7(1)$ extracted
from the local-smeared correlator via Eq.~\ref{two exponential fit}. 
The pentaquark masses at the
chiral limit given in Table~\ref{table:mass_smeared}
are obtained  by linearly extrapolating the results
determined on the set of  five $\kappa_l$ values using Eq.~\ref{chiral}. 
The corresponding results
 for the kaon and nucleon masses at the chiral limit are given in 
Table~\ref{table:parameters}. 
We show all 
chiral extrapolations  in Fig.~\ref{fig:chiral32}.
Note that the slope in the case of  pentaquarks is steeper than
for the nucleon or the kaon which means that the mass gap between
pentaquarks and KN  increases with the quark mass, giving
at $\kappa_l=0.153$ a gap of about 170~MeV as mentioned earlier.
From the values obtained at the chiral limit we can evaluate
 the ratios of the mass of the candidate pentaquark
 in the positive and negative
 parity channels to
the mass of the kaon-nucleon system, $m_{KN}$. The values that we find
for  these ratios are
$1.12 \pm 0.04$ for the negative parity and $1.65 \pm 0.09$ for the positive. 
 Substituting  the physical kaon and
nucleon mass values leads to the determination of  the  mass in physical units.
We find in the negative
and  positive parity channels  the values 
\be
m^-=1.605 \pm 0.058\>{\rm GeV} \hspace*{0.5cm}
m^{+} = 2.36 \pm 0.13\>{\rm GeV}
\ee
respectively.
As we have already pointed out
the ratio between the mass of the $\phi$ meson
evaluated at $\kappa_s=0.155$ and the
mass of the nucleon at the chiral limit is $1.002 \pm 0.025$, 
within  10\% of 
the value of $1.087$ obtained using the physical $\phi$ meson
 and nucleon masses. Similarly the ratio $m_K/m_N=0.493(17)$ at the chiral
limit is very close to the physical value of $0.526$.

\begin{figure}
\epsfxsize=8truecm
\epsfysize=6.truecm
\mbox{\epsfbox{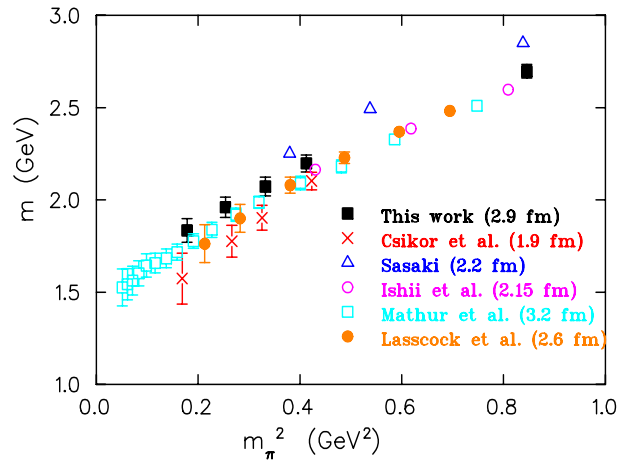}}
\caption{The mass of the lowest eigenstate of the pentaquark system in 
the negative parity channel 
as a function of the pion mass squared.
Filled squares show the results of this work, crosses
are data from Ref.~\cite{Wuppertal}, open triangles from Ref. ~\cite{Sasaki},
open circles 
from Ref.~\cite{Ishii}, open squares  from 
Ref.~\cite{Mathur} and  filled circles from Ref.~\cite{Adelaide}.
In each case we give the spatial length of the lattice.}
\label{fig:results_all}
\end{figure}

Finally we compare our analysis to other lattice studies 
of the pentaquark system. 
All lattice studies so far are carried out in quenched QCD. The
first pioneering studies with
Wilson fermions were carried out by Csikor {\it et al.}~\cite{Wuppertal} using 
a variant of the KN interpolating field, ${\cal J}_{KN}^\prime$ 
where the color index of the $\bar{s}$ is coupled to 
a light quark in the nucleon instead of in the kaon and S. Sasaki~\cite{Sasaki} using the
diquark-diquark interpolating field.
Both found  evidence for a pentaquark state. Motivated by these studies we
looked at the pentaquark density-density correlator in order to learn
about the internal structure of the $\Theta^+$~\cite{latt04}.
Chiu and Hsied ~\cite{Chiu} using overlap fermions performed
a variational analysis using interpolating fields 
${\cal J}_{KN}$,  ${\cal J}_{KN}^\prime$ and
${\cal J}_{DD}$. Their conclusion was that there is a resonance
in  the positive parity channel, which in
  the chiral limit
 has a mass
close to that of the $\Theta^+$(1540). 
The next lattice group to have results on the pentaquark system was
Mathur {\it et al.}~\cite{Mathur}. They used  interpolating fields
 ${\cal J}_{KN}$ and ${\cal J}_{KN}^\prime$~\cite{Wuppertal}.
By studying the scaling of spectral weights
on two volume of  size $12^3\times 28 $ and $16^3\times 28 $ 
using a sequential 
Bayesian method they concluded that
the states that they observed were KN scattering states.
 In the work of  Ishii {\it et al.} in 
addition to periodic 
boundary conditions, antiperiodic boundary conditions 
were used in the spatial direction for the light quarks. With
antiperiodic boundary conditions 
the lowest momentum allowed for each quark is $\pi/L$.
Therefore  in the negative parity channel
switching from periodic to antiperiodic b.c. should
increase the energy of the lowest KN scattering state.
They indeed observed the expected shift in energy concluding
that the lowest state is a KN scattering state.
Our analysis is closest to the analysis
carried out in ref.~\cite{japan}. Diagonalizing 
the $2\times 2$ correlation matrix constructed using ${\cal J}_{KN}$
and ${\cal J}^\prime_{KN}$, and using a large number
of configurations they were able to 
accurately determine the two lowest eigenvalues
in the negative parity channel and check 
scaling of their spectral weights.
Based on this scaling they concluded that
the lowest state is the s-wave KN scattering
state whereas the second is a resonance state
that they identified as the $\Theta^+$. In the positive
 parity channel they obtained only one state  
 for which they concluded  
that it is not the p-wave KN
scattering state. Finally after the completion of our work
a study of the binding energy in the pentaquark system 
as a function of the light quark mass was carried out in Ref.~\cite{Adelaide}.
The fact that they saw no increase in the binding as the
light quark mass decreases led them to conclude that
there is no pentaquark resonance. 
Following  Ref.~\cite{Adelaide} we  compare  our results
given in Table~\ref{table:mass_smeared} for the negative parity channel
to the results from other
lattice groups in Fig.~\ref{fig:results_all}.
There are small variations in the
measured data among different groups
which, at least partly, can be attributed to small
differences  in the strange quark mass. Errors due
to the choice of
the lattice scale $a$ are also not included.

\section{Conclusions}

We have presented a computation of the mass of the pentaquark system
using interpolating fields, which are  motivated by
 the diquark-diquark  and diquark-triquark description proposed for the
internal
structure of the $\Theta^+$.
Despite the difference in the structure of these interpolating fields
the values obtained for the lowest mass of the pentaquark system
 using either interpolating field
are in agreement with each other. They are also in agreement with
the lowest energy eigenvalue determined  from the analysis 
of  the correlation matrices constructed
using  either local or smeared KN and diquark-diquark interpolating fields 
as a basis. 
We  check our lattice techniques  
by studying  the two-pion system in the I=2 channel
where no low lying resonances are present. By analyzing the correlation
matrix constructed using products of pion and rho-meson interpolating
fields we accurately determined
that the two lowest states are the s-wave two-pion and
two-rho scattering states with a mass gap which is about
 320 MeV at a quark mass that
corresponds to pion mass of
about 830 MeV. 
It is explicitly verified that these are s-wave states by
projecting to zero relative momentum on our smallest lattice.
We note
that the energies obtained by diagonalizing the correlation matrix
in the center of mass frame
 are a mixture of scattering
states of  different relative momentum
even when the two lowest states are scattering states
and only at large time separations they yield the correct ground state. 
This is a relevant result
since in many lattices studies it is assumed that a similar  diagonalization 
resolves the scattering states with different relative momentum.
By studying the spectral weights for the
two lowest eigenstates in the two-pion system on our three
lattice volumes we show that the correct  
scaling with the spatial volume sets in at large times requiring accurate 
determination of local correlators.
The spectral weights in the pentaquark system show no volume dependence 
in the  time range where in the two pion system we saw a clear
scaling of spectral weights. Therefore based on our spectral weights results
we can not exclude a resonance pentaquark state. 
In the negative parity channel
our correlation matrix analysis gives accurately one state close to 
the KN threshold with a first excited state that can  be determined
only at very
short time separations.
The mass of the lowest state is consistent 
with the mass extracted from just ${\cal J}_{DD}$ and
${\cal J}_{KN}$. 
We would like to stress that 
the fact that within the time extent of our smaller lattices
the spectral weights do not scale with the volume only leads to the conclusion
that we do not have a single scattering state. Given the fact that within 
our statistics and time ranges we are unable to accurately 
resolve a lower KN- scattering state and a higher close-by single 
particle state
does not permit us to draw any definite conclusion regarding the
existence of the $\Theta^+$. However we have shown in this work
that one would need very accurate data, better interpolating fields and
lattices with large time extension in order to reliably resolve the
low lying states and 
perform a  spectral weights analysis to exclude or show
the existence of the $\Theta^+$.
Given that the lowest eigenstate from the correlation
matrix analysis yields an energy in agreement
with that extracted with ${\cal J}_{DD}$ means that we can use
 the local-smeared 
diquark-diquark interpolating field to compute 
the mass of this state. This is carried out on a set of five values 
of  light quark masses on our large lattice. 
A linear extrapolation to the chiral limit  leads to 
a value for the mass which is
 about 10\% above the KN threshold.  In the positive parity channel 
the two lowest eigenstates are separated by an energy gap of
about 100~MeV 
at the two heaviest pion masses.
 However the mass that we find in the chiral limit,
  determined in the same way as in the
negative parity channel
is about 
920 MeV above the KN threshold and therefore  too high to
 be identified with the $\Theta^+$(1540). 
In summary,  we have shown that within this analysis
in quenched lattice QCD and using
Wilson fermions,
we cannot exclude  a pentaquark state, which in the negative parity
channel has a mass 
about 750~MeV  lower than in the positive channel. However,  
with the lattice sizes used and within our statistics, the existence of
the $\Theta+$ has also not been established from this study. 
In order to reach a definite conclusion regarding
its existence  one would  require a
 more detail and accurate computation involving a larger and better
basis of interpolating fields, lattices with larger temporal
extent and more statistics.

{\bf Acknowledgments:} A. Tsapalis acknowledges funding from the Levendis Foundation.


\begin{thebibliography}{99}
\bibitem{LEPS}LEPS collaboration, T. Nakano     {\it et al.}, Phys. Rev.  Lett.  {\bf 91},   012002 (2003).
\bibitem{experiment}
DIANA collaboration, V. V. Barmin {\it et al.}, Phys. Atom. Nucl.  {\bf 66},   1715   (2003);
CLAS collaboration, S. Stepanyan  {\it et al.}, Phys. Rev.  Lett.  {\bf 91},   252001 (2003); 
SAPHIR collaboration, J. Barth    {\it et al.}, Phys. Lett.        {\bf B572}, 127    (2003).
\bibitem{positive} A.E. Asratyan, A.G. Dolgolkenko and M.A. Kubantsev, Phys. Atom. Nucl. {\bf 67}, 682 (2004);
V. Kubarovsky {\it et al.} (CLAS), Phys. Rev. Lett. {\bf 92}:032001 (2004);
A. Airapetian {\it et al.} (HERMES), Phys. Lett. {\bf B585} (2004) 213;
The ZEUS collaboration, Phys. Lett. {\bf B591} (2004) 7;
M. Abdel-Barv, {\it et al.} (COSY-TOF), Phys. Lett. {\bf B595}, 127 (2004);
A. Aleev {\it et al.} (SVD), submitted to Yad. Fiz.; hep-ex/0401024.
\bibitem{negative} J.Z. Bai {\it et al.} (BES),Phys. Rev. D {\bf 70}:012004 (2004);
The BaBar Collaboration, hep-ex/0408064 ;
The Belle Collaboration, hep-ex/0409010 ;
S.R. Armstrong, hep-ex/0410080; S. Schael {\it et al.} (ALEPH), Phys. Lett. B {\bf 599}, 1 (2004) ;
I. Abt {\it et al.} (HERA-B), Phys. Rev. Lett. {\bf 93}:212003 (2003);
Yu.M. Antipov {\it et al.} (SPHINX), Eur. Phys. J. {\bf A21}, 455 (2004);
M.J. Longo {\it et al.} (HyperCP), Phys. Rev. D {\bf 70}:111101 (2004); 
D.O. Litvintsev (CDF), hep-ex/0410024;
K. Stenson {\it et al.} (FOCUS), hep-ex/0412021;
R. Mizuk {\it et al.} (Belle), hep-ex/0411005;
C. Pinkerton {\it et al.} (PHENIX), J. Phys. G {\bf 30}:S1201 (2004).
\bibitem{negative2} After the submission of this work to Phys. Rev. D the CLAS
collaboration has reported a negative result, M. Battaglier {\it et al.}, 
hep-ex/0510061. 
\bibitem{soliton} D. Diakonov, V. Petrov and M. Polyakov, Z. Phys. A {\bf 359}, 305 (1997). 
\bibitem{nussinov} A. Casher and S. Nussinov, Phys. Lett. {\bf B578}, 124 (2004).
\bibitem{Lipkin} M. Karliner and H. J. Lipkin, Phys. Lett. {\bf B575}, 249 (2003).
\bibitem{diamond}  X.-C. Song and S.-L. Zhu, hep-ph/0403093.
\bibitem{AK} C. Alexandrou and G. Koutsou, Phys. Rev. D {\bf 71}, 014504 (2005).
\bibitem{Jaffe} R. Jaffe and F. Wilczek, Phys. Rev. Lett. {\bf 91}, 232003 (2003).
\bibitem{Wuppertal} F. Csikor, Z. Fodor, S. D. Katz, T.G. Kovacs JHEP 0311, 
070 (2003).
\bibitem{Chiu} T.-W. Chiu, T.-H. Hsieh hep-ph/0403020. 
\bibitem{Sasaki} S. Sasaki, Phys. Rev. Lett. {\bf 93} 152001(2004).
\bibitem{latt04} C. Alexandrou, G. Koutsou and A. Tsapalis, Nucl. Phys. 
(Proc. Suppl.) {\bf 140}, 275 (2005),  hep-lat/0409065.
\bibitem{japan} T. T. Takahashi, T. Umeda, T. Onogi and T. Kunihiro, 
hep-lat/0410025.
\bibitem{Mathur} N. Mathur {\it et al.}, Phys. Rev. D {\bf 70}, 074508 (2004).
\bibitem{Ishii} 
  N. Ishii, {\it et al.},Phys. Rev. D {\bf 71}, 034001 (2005).
\bibitem{Adelaide} B. G. Lasscock, {\it et al.}, hep-lat/0503008.
\bibitem{summary} S. Sasaki, Nucl. Phys. (Proc. Suppl.) {\bf 140}, 127 (2005)
,hep-lat/0410016; G. Fleming, hep-lat/050101. 
\bibitem{pentaq} F. Okiharu, H. Suganuma and T. T. Takahashi, hep-lat/0407007.
\bibitem{Luscher} M. L\"uscher, Nucl. Phys. {\bf B364}, 237 (1991).
\bibitem{Sasaki2} S. Sasaki, T. Blum and S. Ohta, Phys. Rev. D {\bf 65}, 074503
(2002).
\bibitem{smear} S. G\"usken, Nucl. Phys. B (proc. Suppl.) 17, 361 (1991); C. Alexandrou, {\it et al.} Nucl. Phys. {\bf B414}, 815 (1994).
\bibitem{variational} N. A. Campbell, A. Huntley and C. Michael, Nucl Phys.
{\bf B306}, 51 (1988); M. L\"uscher and U. Wolff, Nucl. Phys. {\bf B339}, 222 (1990);
 M. Guagnelli, R. Sommer and H. Wittig, Nucl. Phys. {\bf B535}, 389 (1998).
\bibitem{Leinweber} D. Leinweber, R. M. Woloshyn and T. Draper, Phys. Rev. D {\bf 43} (1991) 1659. 
\bibitem{connection} NERSC archive, G. Kilcup {\it et al.}, hep-lat/9609006.
\bibitem{Luscher scatter} M. L\"uscher, Nucl. Phys. {\bf B354}  (1991) 534.
\bibitem{Wiese} U.-J. Wiese, Nucl. Phys. (Proc. Suppl.) 9 (1989) 609. 
\bibitem{Zhu} S.-L. Zhu, Phys. Rev. Lett. {\bf 91}, 232002 (2003).
\end{thebibliography}
\end{document}